%% file: beaten.tex
\documentclass[12pt]{article}
\input{psfig}
\usepackage{epsfig}
\usepackage{pstricks}
\usepackage{axodraw}
\usepackage{amsmath, amssymb}

 \usepackage{pst-coil}
\def\a{\alpha}
\def\b{\beta}

\def\d{\delta}
\def\e{\epsilon}
\def\f{\phi}
\def\g{\gamma}
\def\h{\eta}

\def\j{\psi}
\def\k{\kappa}
\def\l{\lambda}
\def\m{\mu}
\def\n{\nu}

\def\p{\pi}

\def\r{\rho}
\def\s{\sigma}
\def\t{\tau}

\def\z{\zeta}
\def\D{\Delta}
\def\F{\Phi}
\def\G{\Gamma}

\def\L{\Lambda}
\def\O{\Omega}

\def\Q{\Theta}
\def\S{\Sigma}

\def\ve{\varepsilon}

\def\cl{{\cal L}}

\def\co{{\cal O}}

\def\slpa{\slash{\pa}}                            
\def\bo{{\raise.15ex\hbox{\large$\Box$}}}               
\def\pa{\partial}                                       
\def\pr{\prod}                                          
\def\face{{\raise.2ex\hbox{$\displaystyle \bigodot$}\mskip-2.2mu \llap {$\ddot
        \smile$}}}                                      
\def\dg{\dagger}                                     
\def\wt#1{\widetilde{#1}}                    
\def\VEV#1{\left\langle #1\right\rangle}        

\def\beqa{\begin{eqnarray}}
\def\eeqa{\end{eqnarray}}
\def\NO{\nonumber}

\def\slash#1{\rlap{\hbox{$\mskip 1 mu /$}}#1}      
\def\wt#1{\widetilde{#1}}                    
\def\Bar#1{\overline{#1}}                       
\def\VEV#1{\left\langle #1\right\rangle}        
\def\leftrightarrowfill{$\mathsurround=0pt \mathord\leftarrow \mkern-6mu
        \cleaders\hbox{$\mkern-2mu \mathord- \mkern-2mu$}\hfill
        \mkern-6mu \mathord\rightarrow$}       
\def\dvec#1{\vbox{\ialign{##\crcr
        \leftrightarrowfill\crcr\noalign{\kern-1pt\nointerlineskip}
        $\hfil\displaystyle{#1}\hfil$\crcr}}}           

\def\pl#1#2#3{Phys.~Lett.~{\bf B {#1}} ({#2}) #3}
\def\np#1#2#3{Nucl.~Phys.~{\bf B {#1}} ({#2}) #3}
\def\prl#1#2#3{Phys.~Rev.~Lett.~{\bf #1} ({#2}) #3}
\def\pr#1#2#3{Phys.~Rev.~{\bf D {#1}} ({#2}) #3}

\def\ap#1#2#3{Ann.~of Phys.~{\bf {#1}} ({#2}) #3}
\def\prep#1#2#3{Phys.~Rep.~{\bf {#1}C} ({#2}) #3}

\def\nc#1#2#3{Nuovo Cim.~{\bf {#1}} ({#2}) #3}

\begin{document}
\title{
{\normalsize
\begin{minipage}{5cm}
DESY 02-032\\
March 2002
\end{minipage}}\hspace{\fill}\mbox{}\\[5ex]
{\Large\bf NEUTRINOS, GRAND UNIFICATION\\ 
and LEPTOGENESIS}\thanks{Lectures given at the 
{\it 2001 European School of High-Energy Physics}, Beatenberg, Switzerland}}
\author{W. Buchm\"uller \\
\vspace{3.0\baselineskip}                                               
{\normalsize\it Deutsches Elektronen-Synchrotron DESY, 22603 Hamburg, Germany}
}        
\date{}
\maketitle
\thispagestyle{empty}
\vspace{1cm}
\begin{abstract}
\noindent
Data on solar and atmospheric neutrinos provide evidence for neutrino masses
and mixings. We review some basic neutrino properties, the status of
neutrino oscillations and implications for grand unified theories, 
including leptogenesis.

\end{abstract}

\newpage
\tableofcontents

\newpage

\section{Neutrinos and Symmetries}

Neutrinos have always played a special role in physics due to their close
connection with fundamental symmetries and conservation laws. 
Historically, this development started with Bohr who argued in 1930 that
\begin{itemize}
\item{\bf energy and momentum}\\
might only be statistically conserved \cite{boh30} in order to explain the
continuous energy spectrum of electrons in nuclear $\b$-decay. As a `desparate
way out', Pauli then suggested the existence of a new neutral particle
\cite{pau30}, the neutrino, as carrier of `missing energy'. The emission 
of neutrinos in $\b$-decays could also explain the continuous energy spectrum.
\item{\bf lepton number}\\
As a massive neutral particle, the neutrino can be equal to its antiparticle
\cite{maj37} and thereby violate lepton number, a possibility with far
reaching consequences. Most theorists are convinced that neutrinos are Majorana
particles although at present there are no experimental hints supporting
this belief.
\item{\bf parity and charge conjugation}\\
In the theory of weak interactions \cite{fer33} chiral transformations 
\cite{sj55} have played a crucial role in identifying the correct Lorentz
structure. The resulting $V-A$ theory \cite{vma58} violates the discrete 
symmetries charge conjugation ($C$) and parity ($P$), but it conserves the 
joint transformation $C\!P$.
\item{\bf $C\!P$ invariance}\\
The modern theory of weak interactions is the standard model \cite{sm61} 
with right-handed neutrinos. In its most general form $C\!P$ invariance and lepton
number are violated which, as we shall see, has important implications
for the cosmological matter-antimatter asymmetry.
\item{\bf $C\!P\!T$ and Lorentz invariance}\\
At present neutrino oscillations are discussed as a tool to probe $C\!P\!T$
and Lorentz invariance \cite{ckx98} whose violation is suggested by some 
modifications of the standard model at very short distances. 
\end{itemize}
It therefore appears that we are almost back to the ideas of Bohr, with the
hope for further surprises ahead.

\section{Some Neutrino Properties}

Neutrino physics is a broad field \cite{fy94} involving nuclear physics, 
particle physics, astrophysics and cosmology. In the following we shall
only be able to mention some of the most important neutrino properties.
Most of the lectures will then be devoted to neutrino oscillations,
the connection with grand unified theories and also the cosmological
matter-antimatter asymmetry.

\subsection{Weak interactions}

A major source of neutrinos and antineutrinos of electron type is nuclear
$\b$-decay,
\beqa
A(Z,N) &\rightarrow & A(Z+1,N-1) + e^- + \overline{\n}_e\;,\NO\\
A(Z,N) &\rightarrow & A(Z-1,N+1) + e^+ + \n_e\;,
\eeqa
which includes in particular neutron decay, 
$n \rightarrow p + e^- + \overline{\n}_e$. Muon neutrinos are produced
in $\p$ decays, $\p^{\pm} \rightarrow \m^{\pm} + \n_\m (\overline{\n}_\m), 
e^{\pm} + \n_e (\overline{\n}_e)$ and muon decays, 
$\m^{\pm} \rightarrow e^{\pm} + \overline{\n}_\m (\n_\m) 
+ \n_e (\overline{\n}_e)$.
All these processes are described by Fermi's theory of weak interactions,
\begin{equation}
L_F = -\frac{G_F}{\sqrt{2}} J^{CC}_\m J^{CC\m\dg}\;,
\end{equation}
where $G_F \simeq 1.166 \times 10^{-5}$~GeV$^{-2}$ \cite{rpp00} is
Fermi's constant.
The charged current $J^{CC}_\m$ has a hadronic and a leptonic part,
\beqa
J^{CC}_\m &=& J^{CC(h)}_\m + J^{CC(l)}_\m \;, \\ 
J^{CC(h)}_\m &=& \overline{p}\g_\m (g_V - g_A\g_5) n 
               + f_\p \partial_\m \p^+ + \ldots\;,\\
J^{CC(l)}_\m &=& \overline{\n}_e\g_\m(1-\g_5)e 
           + \overline{\n}_\m\g_\m(1-\g_5)\m + \ldots\;.
\eeqa
Here $f_\p$ is the pion decay constant, and $g_V = 0.98$ and $g_A = 1.22$
are the vector and axial-vector couplings of the nucleon.

An impressive direct test of the $V-A$ structure of the charged weak current
is the helicity suppression in $\p$ decays. Since pions have spin zero,
angular momentum conservation requires the same helicities for the outgoing
antineutrino and charged lepton. On the other hand, the $V-A$ current
couples particles and antiparticles of opposite helicities.
Hence, one helicity flip is needed,
and the decay amplitude is proportional to the mass of the charged lepton.
This immediately yields for the ratio of the partial decay widths,
\begin{equation}
R = \frac{\G(\p \rightarrow e + \bar{\n}_e)}{\G(\p \rightarrow \m + \bar{\n}_\m)}
  = \left(\frac{m_e}{m_\m}\right)^2 
  \left(\frac{m_\p^2 - m_e^2}{m_\p^2 - m_\m^2}\right)^2 = 1.23\times 10^{-4}\;,
\end{equation}
in remarkable agreement with experiment \cite{rpp00},
\begin{equation}
R_{exp} = (1.230 \pm 0.004)\times 10^{-4}\;.
\end{equation} 

From the measurement of the invisible width of the $Z$-boson we know that 
there are three light neutrinos \cite{rpp00}, 
\begin{equation}
N_\n = {\G_{inv}\over \G_{\n\overline{\n}}} = 2.984\pm 0.008\;.
\end{equation}
This is consistent with the existence of $\n_e$, $\n_\m$ and $\n_\t$, i.e. 
one neutrino for each quark-lepton generation. Direct evidence for the 
$\t$-flavour of the third neutrino has recently been obtained by 
the DONUT collaboration
at Fermilab, which observed $\t$-appearance in nuclear emulsions \cite{don01},
\begin{equation}
\n_\t + N \rightarrow \t + X \;.
\end{equation} 
The same reaction will be used to prove 
$\n_\m \rightarrow \n_\t$ oscillations by $\t$-appearance.

\subsection{Bounds on neutrino masses}

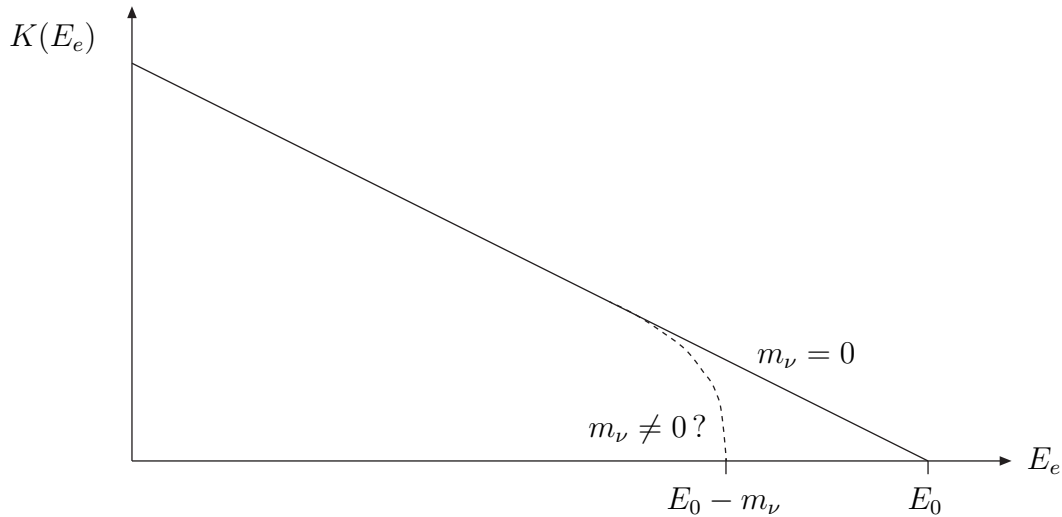
\begin{figure}
\begin{center}
\input{kurie.tex}
\end{center}
\vspace{-0.7cm}
\caption{\label{fig:kurie}\it Kurie plot for tritium beta-decay.}
\end{figure}

Neutrinos are expected to have mass, like all other leptons and quarks. Direct
kinematic limits for tau- and muon-neutrinos have been obtained from the decays
of $\t$-leptons and $\p$-mesons, respectively. The present upper bounds
are \cite{rpp00},
\begin{equation}
m_{\n_\t} < 18.2\ \mbox{MeV}\ (\mbox{95\% CL})\;,\quad 
m_{\n_\m} < 170\ \mbox{keV}\ (\mbox{90\% CL})\;.
\end{equation}
These are bounds on combinations of neutrino masses, 
which depend on the neutrino mixing.
The study of the electron energy spectrum in tritium $\b$-decay over many
years has led to an impressive bound for the electron-neutrino mass.
The strongest upper bound has been obtained by the Mainz collaboration
\cite{mai01}
\begin{equation}\label{mai}
m_{\n_e} < 2.2\ \mbox{eV}\ (\mbox{95\% CL})\;.
\end{equation}
It is based on the analysis of the Kurie plot (fig.~\ref{fig:kurie}) where
the electron energy spectrum is studied near the maximal energy $E_0$,
\begin{equation}
K(E_e) \propto \sqrt{(E_0 - E_e)((E_0 - E_e)^2 - m_\n^2)^{1/2}} \;.
\end{equation}
In the future the bound (\ref{mai}) is expected to be improved to 0.3~eV 
\cite{kat01}.   

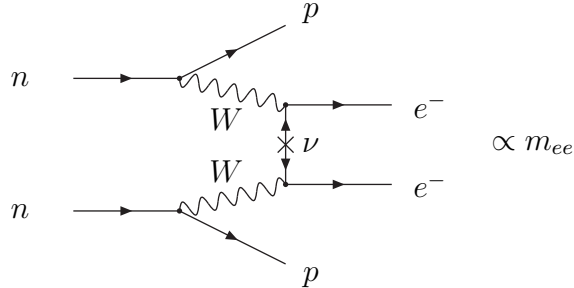
\begin{figure}
\begin{center}
\input{doubb.tex}
\end{center}
\vspace{-0.7cm}
\caption{\label{fig:doubb}\it Effective electron mass in $0\n2\b$-decay.}
\end{figure}

The most stringent bound so far has been obtained for the effective
electron-neutrino mass $m_{ee}$ in neutrinoless double $\b$-decay 
(fig.~\ref{fig:doubb}),
\begin{equation}
A(Z,N) \rightarrow A(Z+2,N-2) + e^- + e^- \;.
\end{equation}
The Heidelberg-Moscow collaboration presently quotes the bound \cite{hei01},
\begin{equation}\label{hei}
|m_{ee}| < 0.38\ \k\ \mbox{eV}\ (\mbox{95\% CL})\;,
\end{equation}
where $\k = \co(1)$ represents the effect of the hadronic matrix element.
The quantity  $m_{ee}$ is a Majorana mass. Hence, a non-zero result would
mean that lepton number violation has been discovered! The GENIUS
project has the ambitious goal to improve the present bound (\ref{hei}) 
by about two orders of magnitude \cite{gen01}.

\subsection{Cosmology and astrophysics}

Big bang cosmology successfully describes the primordial abundances of
the light elements $^4$He, D and $^7$Li. This also leads to predictions for
the baryon density $\O_B$ and the effective number of light neutrinos at
the time of neutrino decoupling ($T \sim 1$~MeV, $t \sim 1$~s) \cite{dib01},
\begin{equation}\label{neff}
1.2 < N^{eff}_\n < 3.3 \ (\mbox{99\% CL})\;.
\end{equation}
Before the precise measurement of the Z-boson width this bound was already
a strong indication for the existence of not more than three weakly
interacting `active' neutrinos. The bound also strongly constrains the
possibility of large lepton asymmetries at the time of nucleosynthesis
and the possible existence of `sterile' neutrinos which do not interact
weakly. Detailed studies show that mixing effects would be sufficient to
bring a sterile neutrino into thermal equilibrium. This would imply
$N^{eff}_\n = 4$, in contradiction to the bound (\ref{neff}). The
existence of a sterile neutrino is therefore disfavoured \cite{dib01}.

The standard cosmological model further predicts the existence of relic
neutrinos with temperature $T \sim 1.9$~K and an average number density 
around 100/cm$^3$ per neutrino species. For neutrino masses between 
$3\times 10^{-2}$~eV (see section 4.3) and 2~eV (cf.~(\ref{mai})) this yields 
the contribution to the cosmological energy density,
\begin{equation}
0.001 < \O_\n h^2 < 0.1 \; ;
\end{equation}
here $\O_\n = \r_\n/\r_c$, where $\r_\n$ is the neutrino energy density and 
$\r_c = 3H_0^2/(3\p G_N)=1.05 h^2 \times 10^4$~eV/cm$^3$ is the critical 
energy density. Hence, $\O_\n$ is substantially smaller than the contributions 
from the `cosmological constant' ($\O_\L$) and from cold dark matter
($\O_{CDM}$), but it may still be as large as the contribution from baryons,
$\O_B h^2 \simeq 0.02$. 

\begin{figure}
\begin{center}
\input{burst.tex}
\end{center}
\vspace{-0.7cm}
\caption{\label{fig:burst}\it Z-burst in the annihilation of ultrahigh
energy cosmic neutrinos and relic neutrinos.}
\end{figure}
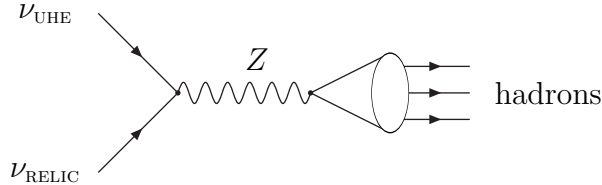

\begin{figure}
\centerline{\epsfig{file=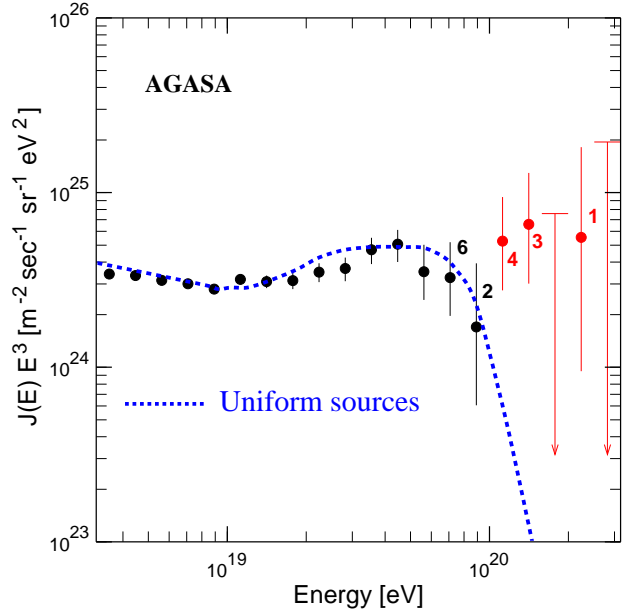,width=9cm}}
\caption{\label{fig:agasa}\it Observed energy spectrum of cosmic rays extending
beyond the GZK cutoff \protect\cite{aga98}.}
\end{figure}
\begin{figure}
\centerline{\epsfig{file=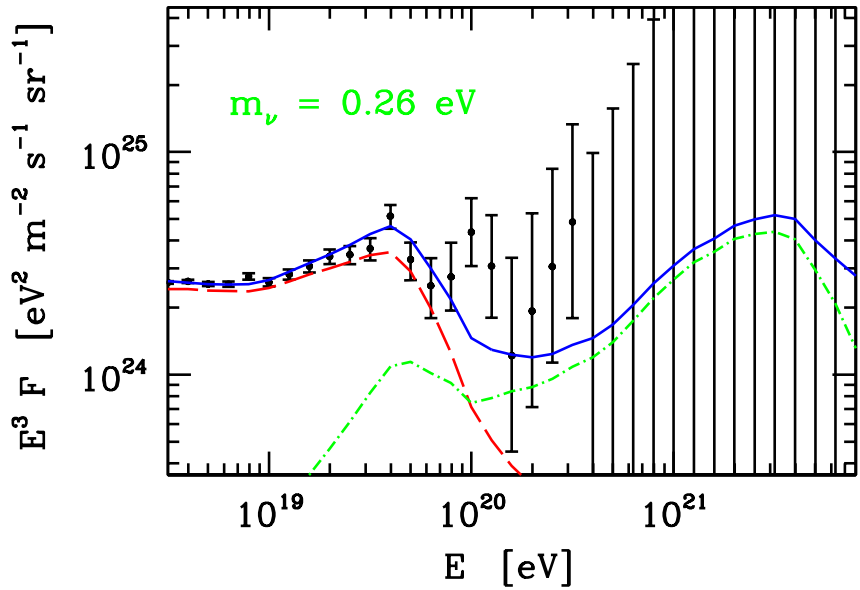,width=12cm}}
\vspace{-1cm}
\caption{{\it Energy spectrum of cosmic rays predicted from Z-bursts of
extragalactic cosmic neutrinos and relic neutrinos with mass} 
$m_\n = 0.26\ \mbox{eV}$ \protect\cite{fkr01}.\label{fig:uhe}}
\end{figure}

The detection of the relic neutrinos is an outstanding problem. 
It has been suggested that this may be possible by means of `Z-bursts'
(fig.~\ref{fig:burst}), which
result from the resonant annihilation of ultrahigh energy cosmic
neutrinos (UHEC$\n$s) with relic neutrinos \cite{wei82,rou93}. On resonance, 
the annihilation cross section is enhanced by several orders of magnitude.
In the rest system of the massive relic neutrinos the resonance condition
is fullfilled for the incident neutrino energy
\begin{equation}
E^{res}_\n = {M_Z^2\over 2m_\n} = 4.2\times 10^{21}\ \mbox{eV}
             \left(m_\n [\mbox{eV}]\right)^{-1}\;,
\end{equation}
where $M_Z$ is the Z-boson mass. For neutrino masses $\co(\mbox{eV})$
these energies are remakably close to the highest energy cosmic rays observed.

Z-bursts are indeed a possible explanation \cite{fms99,wei99} of the observed 
excess at
energies above the Greisen-Zatsepin-Kuzmin (GZK) cutoff 
$E_{GZK} = 4\times 10^{19}$~GeV (fig.~\ref{fig:agasa}). In a detailed
study of extragalactic neutrinos a quantitive description of the data
(fig.~\ref{fig:uhe}) is obtained for neutrino masses \cite{fkr01},
\begin{equation}
m_\n = 0.26^{+0.20}_{-0.14}\ \mbox{eV}\;.
\end{equation}
In fig.~\ref{fig:uhe} the first bump at $E=4\times 10^{19}$~GeV represents
protons produced at higher energies and accumulated just above the GZK cutoff.
The second bum at $E=3\times 10^{21}$~eV  reflects the Z-burst.

A confirmation of the Z-burst hypothesis would prove the existence of
relic neutrinos and at the same time provide an absolute neutrino mass
measurement!

\section{Weyl-, Dirac- and Majorana-Neutrinos}

\subsection{C, P and CP}

The standard model is a chiral gauge theory, i.e. left- and right-handed
fields have different electroweak interactions. Left-handed quarks and
leptons are doublets with respect to the gauge group SU(2) of the weak 
interactions, whereas right-handed quarks and leptons are singlets,
\begin{equation}
q_L = \left(\begin{array}{c} u_L\\ d_L \end{array}\right)\;, 
\quad u_R\;, \quad d_R\;,
\quad l_L = \left(\begin{array}{c} \n_L\\ e_L \end{array}\right)\;, 
\quad e_R\;.
\end{equation}
For massless particles helicity and
chirality are identical. The chirality transformation corresponds to a
multiplication of the Dirac spinor by $\g_5$,
\begin{equation}
\g_5 \j_L(t,\vec{x}) = - \j_L(t,\vec{x})\;, 
\quad \g_5 \j_R(t,\vec{x}) = + \j_R(t,\vec{x})\;.
\end{equation}
Since left- and right-handed fields, which have different gauge interactions,
also have opposite chiralities, the standard model is called a chiral 
gauge theory. 

For the standard model, the transformation properties of the fields under 
$C$, $P$ and $C\!P$ are of fundamental 
importance. Charge conjugation and parity are defined as
\beqa
C: \qquad \j(t,\vec{x}) \rightarrow 
   \j^C(t,\vec{x}) &=& C \overline{\j}^T\!(t,\vec{x})\;,\\
P: \qquad \j(t,\vec{x}) \rightarrow 
   \j^P(t,\vec{x}) &=& P \j(t,-\vec{x})\;.
\eeqa
Here $C=i\g_2\g_0$, $\overline{\j} = \j^\dg \g_0$ and $P=\g_0$.
For simplicity, we have set possible phase factors in the definition
of $C$ and $P$ equal to 1.
One easily verifies that $C$ and $P$ transformations change chirality,
\beqa
\g_5 \j^P_L(t,\vec{x}) = 
   \g_5 \g_0 \j_L(t,-\vec{x}) = + \j_L^P(t,\vec{x})\;,\\
\g_5 \j^C_L(t,\vec{x}) = 
   \g_5 i\g_2 \j^*_L(t,\vec{x}) = + \j_L^C(t,\vec{x})\;.
\eeqa
Hence, parity and charge conjugated left-handed fermions are right-handed
and vice versa.

In general, a Dirac fermion can be decomposed into left- and right-handed
parts,
\begin{equation}
\j = {1+\g_5\over 2}\j + {1-\g_5\over 2}\j = \j_R + \j_L \;.
\end{equation}
Parity and charge conjugation then relate these two components. However, in
the standard model left- and right-handed fields have different weak
interactions. For the up-quark, for instance, one has, 
\begin{equation}
u = u_R + u_L\;, \quad   u_L = q_L^1\;,  
\end{equation}
i.e. $u_L$, the left-handed partner of $u_R$, changes under SU(2) gauge 
transformations. Hence, one cannot define $P$ and $C$ transformations 
which commute with the 
electroweak gauge symmetries. For neutrinos the situation is even worse,
since right-handed neutrinos don't even exist
in the minimal standard model! Only the $C\!P$ transformation is well defined,
\begin{equation}
C\!P: \qquad \j(t,\vec{x}) \rightarrow 
   \j^{CP}\!(t,\vec{x}) = C\!P \overline{\j}^T\!(t,-\vec{x})\;,
\end{equation}
since it does not change chirality,
\begin{equation}
\g_5 \j^{CP}_L = - \j^{CP}_L\;, \quad \g_5 \j^{CP}_R = + \j^{CP}_R\;.
\end{equation}
Hence, in principle, $C\!P$ could be a symmetry of the standard model.

Neutrinos are the quanta of the left-handed field $\n_L(x)$. Neutrino states
are created by the field operator which has the mode expansion,
\begin{equation}\label{nleft}
\n_L(x) = \int d\bar{p}
\left(b(p)u_L(p)e^{-ip\cdot x} + d^\dg(p)v_L(p)e^{ip\cdot x}\right)\;,
\end{equation}
where $d\bar{p} = d^3p/[(2\p)^3 2E]$.
The operators $b^\dg(p)$ and $d^\dg(p)$ create from the vacuum neutrino and
antineutrino states, respectively,
\begin{equation}
|\n;p\rangle = b^\dg(p) |0\rangle\;, \quad
|\bar{\n};p\rangle = d^\dg(p) |0\rangle\ \;.
\end{equation}
The operators for energy, momentum, helicity and lepton number are given
by (cf.~\cite{iz80}),
\beqa
P_\m &=& {i\over 2} \int d^3x :\j^\dg \dvec {\pa_\m} \j:\
=\int d\bar{p}\ p_\m \left(b^\dg(p)b(p)+d^\dg(p)d(p)\right)\;,
\label{opp}\\
\vec{\S}\cdot \vec{P} &=& {i\over 2} \int d^3x :\j^\dg \s^i\dvec {\pa^i} \j:\
= - \int d\bar{p}\ E \left(b^\dg(p)b(p)-d^\dg(p)d(p)\right)\;,
\label{ops}\\
L &=& \int d^3x :\j^\dg \j:\
= \int d\bar{p} \left(b^\dg(p)b(p)-d^\dg(p)d(p)\right)\;;
\label{opl}
\eeqa
here $\s^i = {i\over 2} \e_{ijk}\g^j\g^k$.
Using the usual anticommutation relations for creation and annihilation 
operators one reads off from eqs.~(\ref{opp})-(\ref{opl}),
\beqa
P_\m |\n;p\rangle = p_\m |\n;p\rangle \;, \quad
{\vec{\S}\cdot\vec{P}\over E} |\n;p\rangle = - |\n;p\rangle 
= - L |\n;p\rangle\;,\\
P_\m |\bar{\n};p\rangle = p_\m |\bar{\n};p\rangle \;, \quad
{\vec{\S}\cdot\vec{P}\over E} |\bar{\n};p\rangle = + |\bar{\n};p\rangle 
= - L |\bar{\n};p\rangle\;,
\eeqa
i.e. the helicity (lepton number) of antineutrinos is positive (negative). 
Note, that helicity and lepton number always have opposite sign.

\subsection{Mass generation in the standard model}

Since the standard model is a chiral gauge theory quark and lepton masses
can only be generated via spontaneous symmetry breaking. Starting from the
Yukawa interactions
\begin{equation}
\cl_Y^{(q,e)} = h_{uij}\overline{q}_{Li}u_{Rj}H_1 +
        h_{dij}\overline{q}_{Li}d_{Rj}H_2 +
        h_{eij}\overline{l}_{Li}e_{Rj}H_2\; + \; h.c.\;,
\end{equation}
the vacuum expectation values of the Higgs fields, $\VEV H_1 = v_1$ and 
$\VEV H_2 = v_2$, lead to the mass terms,
\begin{equation}
\cl_M^{(q,e)} = m_{uij}\overline{u}_{Li}u_{Rj} +
        m_{dij}\overline{d}_{Li}d_{Rj} +
        m_{eij}\overline{e}_{Li}e_{Rj}\; + \; h.c.\;,
\end{equation}
with the mass matrices for up-quarks, down-quarks and charged leptons,
\begin{equation}
m_u = h_u v_1\;, \quad m_d = h_d v_2\;, \quad m_e = h_e v_2\;.
\end{equation} 
The mass matrices are diagonalized by bi-unitary transformations,
\begin{equation}
V^{(u)\dg}m_u\widetilde{V}^{(u)} = m_u^{diag}\;,\quad
V^{(d)\dg}m_d\widetilde{V}^{(d)} = m_d^{diag}\;,\quad
V^{(e)\dg}m_e\widetilde{V}^{(e)} = m_e^{diag}\;,\quad
\end{equation}
with $V^{(u)\dg}V^{(u)} = 1$, etc. , which also define the transition
from {\it weak} eigenstates $\j_{L\a}$, $\j_{R\a}$ to {\it mass} eigenstates
$\j_{L i}$, $\j_{R i}$,
\begin{equation}
u_{L\a} = V^{(u)}_{\a i}u_{Li}\;,\quad d_{L\a} = V^{(d)}_{\a i}d_{Li}\;,
\ldots , e_{R\a} = \widetilde{V}^{(e)}_{\a i}e_{Ri}\;.
\end{equation}

Since in general the transformation matrices are different for up- and
down-quarks one obtains a mixing between mass eigenstates in the charged
current (CC) weak interactions,
\beqa
{\cal L}^{(q)}_{EW} &=& -{g\over\sqrt{2}}\sum_{\a}
\overline{u}_{L\a} \g^\m d_{L\a}\ W^+_\m \ + \ldots \NO\\ 
&=& -{g\over\sqrt{2}}\sum_{i,j}
\overline{u}_{L i} \g^\m  V_{ij} d_{Lj}\ W^+_\m \ + \ldots\;,
\eeqa
where $V_{ij} = V^{(u)\dg}_{i\a}V^{(d)}_{\a j}$ is the familiar CKM mixing
matrix. In general the Yukawa couplings are complex; hence, also the
CKM matrix is complex, which leads to $C\!P$ violation in weak interactions.

Without right-handed neutrinos, weak and mass eigenstates can always be
chosen to coincide for leptons. As a consequence, there is no mixing in
the leptonic charged current,
\begin{equation}
{\cal L}^{(l)}_{EW} = -{g\over\sqrt{2}}\sum_{\a}
\overline{e}_{L i} \g^\m \n_{L i}\ W^-_\m \ + \ldots \;,
\end{equation}
and electron-, muon- and tau-number are separately conserved.
On the contrary, in the
quark sector only the total baryon number is conserved. 

\subsection{Neutrino masses and mixings}

The simplest, and theoretically favoured, way to introduce neutrino masses
makes use of right-handed neutrinos $\n_R$. This allows additional
Yukawa couplings and a Majorana mass term,
\begin{equation}
\cl_Y^{(\n)} = h_{\n ij}\overline{l}_{Li}\n_{Rj}H_1 +
               {1\over 2} M_{ij} \overline{\n_R^c}_i\n_{Rj}\; + \; h.c.
\end{equation}
The Yukawa interaction defines the quantum numbers of the right-handed
neutrino: it carries lepton number, which is a global charge, but no colour,
weak isospin or hypercharge, which are gauge quantum numbers. Hence,
a Majorana mass term is allowed for the right-handed neutrinos, consistent
with the gauge symmetries of the theory. It is very important that these
masses are {\it not} generated by the Higgs mechanism and can therefore be
much larger than ordinay quark and lepton masses. This leads to light neutrino
masses via the seesaw mechanism \cite{yan79}.

The Yukawa interaction which couples left-handed and right-handed neutrinos
yields after spontaneous symmetry breaking the Dirac neutrino mass matrix
$m_D = h_{\n}v_1$, so that the complete mass terms are given by
\begin{equation}
\cl_M^{(\n)} = m_{Dij} \overline{\n_L}_i \n_{Rj} +
             {1\over 2} M_{ij} \overline{\n_R^c}_i\n_{Rj}\; + \; h.c.
\end{equation}
In order to obtain the mass eigenstates one has to perform a unitary
transformation. Using
$\overline{\n_L}m_D\n_{R}=\overline{\n_R^c}m_D^T\n_{L}^c$, the mass terms
can be written in matrix form
\begin{equation}
\cl_M^{(\n)} = {1\over 2} 
\left(\begin{array}{cc}\overline{\n_L} &\overline{\n_R^c}\end{array}\right)
\left(\begin{array}{cc} 0 & m_D \\ m_D^T & M \end{array}\right)
\left(\begin{array}{c} \n_L^c \\ \n_R \end{array}\right)\; + \; h.c.
\end{equation}
The unitary matrix which diagonalizes this mass matrix is easily constructed
as power series in $\xi=m_D/M$. Up to terms $\co (\xi^3)$ one obtains
\begin{equation}
\left(\begin{array}{c} \n_L   \\ \n_R^c \end{array}\right) =
\left(\begin{array}{cc} 1 - {1\over 2}\xi\xi^\dg & \xi \\
      -\xi^\dg & 1 - {1\over 2}\xi\xi^\dg \end{array}\right) 
\left(\begin{array}{c} L   \\ R^c \end{array}\right) \;.
\end{equation}
In terms of the new left- and right-handed fields, $L$ and $R$, 
the mass matrix is diagonal,
\begin{equation}
\cl_M^{(\n)} = {1\over 2} 
\left(\begin{array}{cc}\overline{L} &\overline{R^c}\end{array}\right)
\left(\begin{array}{cc} m_\n & 0 \\ 0 & M \end{array}\right)
\left(\begin{array}{c} L^c \\ R \end{array}\right)\; + \; h.c.
\end{equation}
Here the mass matrix $m_\n$ is given by
\begin{equation}\label{wettss}
m_\n = - m_D {1\over M} m_D^T\;.
\end{equation}
This is the famous seesaw mass relation \cite{wet81}. 
With $M \gg m_D$, one obviously has 
$m_\n \ll m_D$. As an example, consider the case of just one generation.
Choosing for $m_D$ the largest know fermion mass, 
$m_D \sim m_t \sim 100$~GeV, and for $M$ the unification scale of unified
theories, $M \sim 10^{15}$~GeV, one finds $m_\n \sim 10^{-2}$~eV, which is
precisely in the range of present experimental indications for neutrino
masses. 

$m_\n$ and $M$ are the mass matrices of the light neutrinos and their heavy
partners, respectively. The corresponding mass eigenstates are Majorana
fermions,
\begin{equation}
\n = L + L^c = \n^c\;, \quad N = R^c + R = N^c\;.
\end{equation}
$L, R^c$ and $L^c, R$ are the corresponding left- and right-handed components,
\begin{equation}
L = {1-\g_5\over 2}\n\;, \quad R^c = {1-\g_5\over 2}N\;, \quad\mbox{etc.}
\end{equation}
In terms of the Majorana fields the mass terms read,
\beqa\
\cl_M^{(\n)} &=& {1\over 2} \overline{\n}m_\n {1+\g_5\over 2}\n +
{1\over 2} \overline{N}m_\n {1+\g_5\over 2}N\; + \; h.c.\NO\\
&=& {1\over 2} \overline{\n}\left(\mbox{Re}\{m_\n\} + 
i \mbox{Im}\{m_\n\}\g_5\right)\n +
{1\over 2} \overline{N}\left(\mbox{Re}\{M\} + i \mbox{Im}\{M\}\g_5\right)N \;.
\label{numass}
\eeqa
Note, that in general the mass matrices have real and imaginary parts.
This is a consequence of complex Yukawa couplings and a possible source
of $C\!P$ violation.

The mode expansion of the Majorana field operator reads,
\begin{equation}
\n(x) = \n^c(x) = \int d\bar{p} \sum_{i=1}^2
\left(a_i(p)u_i(p)e^{-ip\cdot x} + a^\dg_i(p)v_i(p)e^{ip\cdot x}\right)\;,
\end{equation}
where $u_i(p)$ and $v_i(p) = u^c_i(p)$ are now solutions of the massive
Dirac equation. In the massless case the connection with the left-handed
neutrino field (\ref{nleft}) is very simple,
\begin{equation}
a_1(p) = b(p)\;, \quad a_2(p) = d(p)\;, \quad u_1(p) = u_L(p)\;, \quad
u_2(p) = v_L^c(p)\;,
\end{equation}
i.e. the two spin states of the Majorana neutrino are identical with
the neutrino and antineutrino states. In the massless case these states
carry different lepton number, which is conserved. For massive Majorana
neutrinos lepton number is violated by the Majorana mass term which
couples the two polarization states.  

So far we have only discussed the simplest version of the seesaw mechanism. 
Additional Higgs fields also lead to a direct mass term for the 
left-handed neutrinos \cite{ms81}. In principle, neutrino masses could also
be generated by radiative corrections. This is a possible alternative to
the seesaw mechanism. Particularly attractive are supersymmetric models with 
broken R-parity \cite{hpx02}, which lead to signatures testable at colliders.

Neutrino masses imply mixing in the leptonic charged current. If the neutrino
mass matrix is diagonalized by the unitary matrix $U^{(\n)}$,
\begin{equation}
U^{(\n)\dg} m_\n U^{(\n)*} = - \left(\begin{array}{ccc} m_1 & 0 & 0 \\
  0 & m_2 & 0 \\ 0 & 0 & m_3 \end{array}\right)\;,
\end{equation}
the mixing matrix $U$ appears in the charged current,
\begin{equation}
{\cal L}^{(l)}_{EW} = -{g\over\sqrt{2}}\sum_{ij}
\overline{e}_{Li} \g^\m U_{ij}\n_{Lj}\ W^-_\m \ + \ldots \;,
\end{equation}
where 
\begin{equation}
U_{ij} = U^{(e)\dg}_{i\a}U^{(\n)}_{\a j}\;.
\end{equation}
$e$-, $\m$- and $\t$-number are now no longer separately conserved. The
mixing matrix is frequently written in the following form,
\begin{equation}
U = \left(\begin{array}{ccc} U_{e1} & U_{e2} & U_{e3} \\ 
U_{\m 1} & U_{\m 2} & U_{\m 3} \\ U_{\t 1} & U_{\t 2} & U_{\t 3}
\end{array}\right)\;,
\end{equation}
and can be expressed in the standard way as product of three rotation in the
12-, 13- and 23-planes,
\begin{equation}
U = \left(\begin{array}{ccc} 1 & 0 & 0 \\ 0 & c_{23} & s_{23} \\
0 & -s_{23} & c_{23} \end{array}\right)
\left(\begin{array}{ccc} c_{13} & 0 & s_{13}e^{-i\d} \\ 0 & 1 & 0 \\
-s_{13}e^{i\d} & 0 & c_{13} \end{array}\right)
\left(\begin{array}{ccc} c_{12} & s_{12} & 0 \\ -s_{12} & c_{12} & 0 \\
0 & 0 & 1 \end{array}\right)\;.
\end{equation}
As we shall see in the next section, present data are consistent with
a small value of $s_{13}$. The $3\times 3$ mixing matrix then becomes a
product of two $2\times 2$ matrices describing  mixing between the first and
second, and the second and third generations, respectively. This very simple
mixing pattern appears to be sufficient to account for the solar and the
atmospheric neutrino deficits.

\section{Neutrino Oscillations}

\begin{figure}
\centerline{\epsfig{file=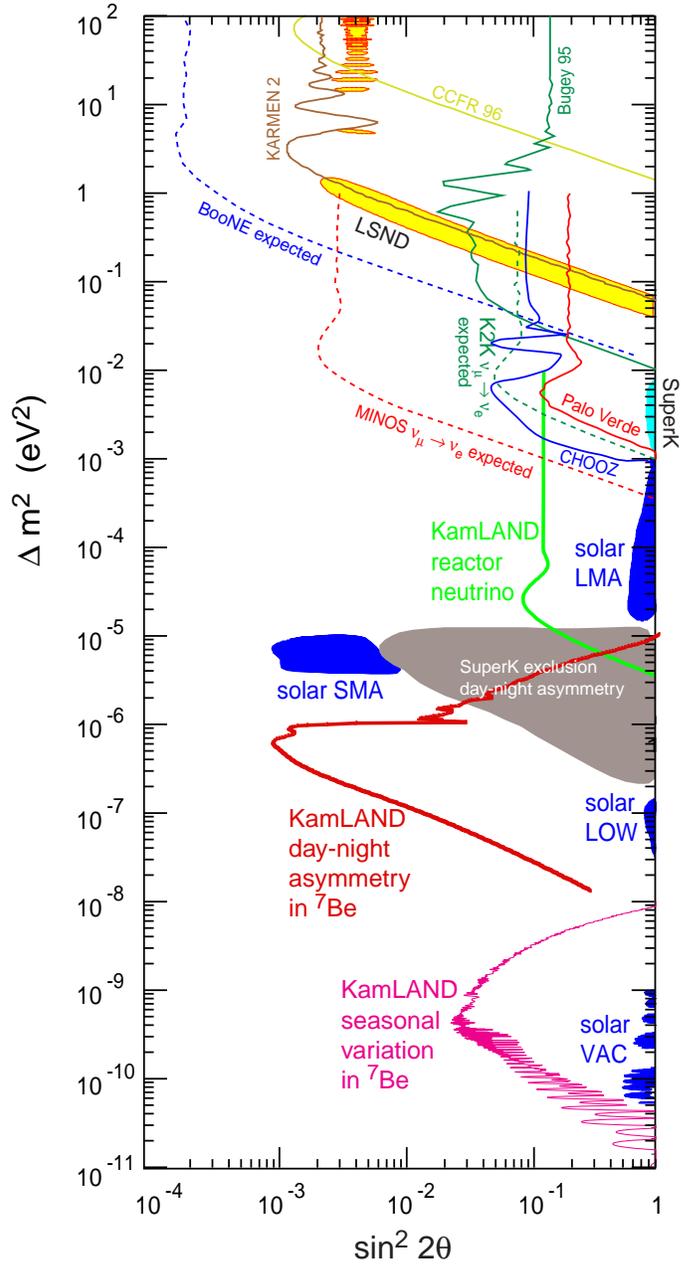,width=10cm}}
\caption{\it The most important exclusion limits as well as preferred 
parameter regions from neutrino oscillation experiments assuming two-flavour
oscillatons \protect\cite{mur00}.
\label{fig_mur}}
\end{figure}

In recent years there has been a wealth of experimental data in neutrino
physics, and we can look forward to important new results also in the
coming years. The present situation is summarized in fig.~\ref{fig_mur} 
which is taken from the review of particle physics. The latest most important
result is the first data from SNO which we shall discuss in section~4.3.
Detailed discussions of neutrino oscillations and references can be
found in the reviews \cite{bgg99}-\cite{ggn02}.

\subsection{Oscillations in vacuum}

Neutrino oscillations \cite{pon57,mns62}
are a very intriguing quantum mechanical effect similar
to the well known oscillations between $K^0$- and $\overline{K^0}$-mesons. 
They occur because of the mixing in the charged weak
current discussed in the previous section. The neutral and charged  current 
weak interactions of neutrinos are described by the lagrangian
\begin{equation}
{\cal L}_{EW} = -\sum_{\a,i}\left\{
{g\over 2\cos{\Theta_W}}\overline{\n}_{L\a} \g^\m  \n_{L\a}\ Z_\m +
{g\over\sqrt{2}}\overline{e}_{L\a} \g^\m  U_{\a i} \n_{Li}\ W^-_\m \ + h.c.
\right\}\;,
\end{equation}
where the fields $e_{L\a}$, $\a = 1\ldots 3$, represent the mass eigenstates
of electron, muon and tau, and the fields $\n_{Li}$, $i = 1\ldots n\geq 3$, 
correspond to neutrino mass eigenstates. Hence, the charged lepton $e_\a$ couples 
to the neutrino flavour eigenstate $\n_\a$, 
which is a linear superposition of mass eigenstates,
\begin{equation}\label{feigen}
\n_\a  = \sum_i U^*_{\a i} \n_i \;.
\end{equation}
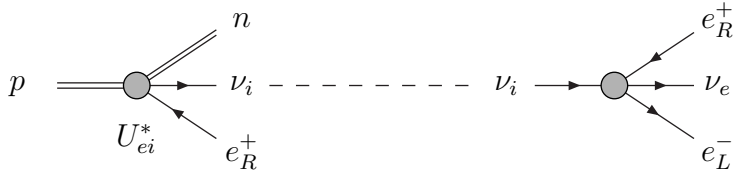
\begin{figure}
\begin{center}
\input{nuosc.tex}
\end{center}
\caption{\label{fig:nuosc}\it Neutrino production, propagation and 
absorption.}
\end{figure}
Here $\n_\a$ and $\n_i$ are spinors in flavour space, and
$U$ is a $n\times n$ unitary matrix. Three linear combinations of mass
eigenstates have weak interactions, and are therefore called {\it active},
whereas $n-3$ linear combinations are {\it sterile}, i.e. they don't feel the
weak force. In the case $n=4$, for instance, the sterile neutrino is given by
\begin{equation}
\n_s  = \sum_i U^*_{4 i} \n_i \;.
\end{equation}
In the following we will restrict ourselves to the case of three active
neutrinos and shall not discuss the positive signature for oscillations from the
LSND experiment \cite{lsnd} at Los Alamos, which would require the existence
of a sterile neutrino.

Neutrinos are relativistic particles whose propagation is described by the
Dirac equation. Consider now a neutrino beam propagating in z-direction
from a source at $z=0$ to a detector at $z=L$. The probability for 
neutrino oscillation is usually derived by considering first the time
evolution of a spatially homogeneous state. Some hand waving arguments are
then needed to obtain the physically relevant probability for oscillations
in space, which, with lack of fortune, can also give the wrong result. 
As pointed out by Stodolsky, all this confusion can be avoided by realizing 
that the system corresponds to a stationary state \cite{sto98,ruj00} which
is completely
characterized by the energy spectrum of the beam. For simplicity, we first
consider the case of fixed energy $E$. The wave packet description of neutrino
oscillations is discussed in ref.~\cite{giu01}. 

Neglecting the mixing between neutrinos and antineutrinos, which is
suppressed by $\left(m_i/E\right)^2$, the hamiltonian of the system reads
in the mass eigenstate basis,
\beqa\label{massham}
H_0 &=& \left(\begin{array}{ccc} \sqrt{\hat{p}^2 + m_1^2} & 0 & 0 \\
0 & \sqrt{\hat{p}^2 + m_2^2} & 0 \\ 0 & 0 & \sqrt{\hat{p}^2 + m_3^2}
\end{array}\right)\NO\\ \NO \\
&\simeq& \left(\begin{array}{ccc} \hat{p} + {m_1^2\over 2\hat{p}} & 0 & 0 \\
0 & \hat{p} + {m_2^2\over 2\hat{p}} & 0 \\ 0 & 0 & 
\hat{p} + {m_3^2\over 2\hat{p}}\end{array}\right)\; ;
\eeqa
here $\hat{p}$ is the momentum operator conjugate to the $z$-coordinate, and
we have assumed that the relevant eigenvalues are much larger than the
neutrino masses $m_i$. To describe the neutrino oscillation 
$\n_\a \rightarrow \n_\b$ we have to find an energy eigenstate,
\begin{equation}
H_0 \n_E^{(\a)}(z) = E \n_E^{(\a)}(z) \;,
\end{equation}
which satisfies the boundary condition
\begin{equation}
\n_E^{(\a)}(0) = \n_\a \;.
\end{equation}
The solution reads
\begin{equation}
\n_E^{(\a)}(z) = \sum_i U^*_{\a i} \n^{(i)}_{p_i}(z) \;,
\end{equation}
where the momentum eigenstates are given by
\begin{equation}
\n^{(i)}_{p_i}(z) = e^{ip_i z}\n_i\; ,
\end{equation}
with $p_i \simeq E - m_i^2/(2E)$. For the dependence of the wave function 
$\n_E^{(\a)}(z)$ on the $z$-coordinate one then obtains,
\begin{equation}
\n_E^{(\a)}(z) = \sum_i U^*_{\a i} e^{i p_i z} \n_i \;.
\end{equation}
Using the unitarity of the mixing matrix, $U^\dg U=1$, this yields for
the flavour transition probability between the source at $z=0$ and the
detector at $z=L$,
\beqa\label{osc1}
P(\n_\a \rightarrow \n_\b) &=& |\n^\dg_\b \n^{(\a)}_E(L)|^2 \NO\\
&=& \d_{\b\a} + \sum_{i,j} U_{\b i}U_{\a i}^*U_{\b j}^*U_{\a i}
    \left(e^{-i\D m^2_{ij}{L\over 2E}}-1\right)\;,
\eeqa
where
\begin{equation}
\D m^2_{ij} = m_i^2 - m_j^2\;.
\end{equation}
Terms with $i=j$ do not contribute in the sum (\ref{osc1}), and a very
useful form of transition probability is 
\beqa\label{master}
P(\n_\a \rightarrow \n_\b) = \d_{\a\b} 
&-& 4 \sum_{i>j} \mbox{Re}\left(U_{\a i}^* U_{\b i} U_{\a j} U_{\b j}^*\right)
             \sin^2{\left(\D m^2_{ij}{L\over 4E}\right)}\NO\\
&+& 2 \sum_{i>j} \mbox{Im}\left(U_{\a i}^* U_{\b i} U_{\a j} U_{\b j}^*\right)
             \sin^2{\left(\D m^2_{ij}{L\over 2E}\right)}\;.
\eeqa
Note the different oscillation lengths of the real and imaginary parts. In
the realistic case of a neutrino beam with some energy spectrum $\r(E)$, one
has to integrate over energy,
\begin{equation}
\bar{P}(\n_\a \rightarrow \n_\b) = \int dE \r(E) P(\n_\a \rightarrow \n_\b)\;.
\end{equation}

All patterns of neutrino oscillations including $C\!P$ violation are described
by the master formula (\ref{master}). Let us discuss some of its properties:
\begin{itemize}
\item
The dependence on $L$ shows the expected oscillatory behaviour, which
is an interference effect and dissapears for  $\D m^2_{ij} =0$.
\item
Due to the unitarity of the mixing matrix $U$ the total flux of neutrinos
is conserved,
\begin{equation}
\sum_{\b} P(\n_\a \rightarrow \n_\b) = \sum_i |U_{\a i}|^2 = 1\;.
\end{equation}
\item
In the simplest case of only 2 flavours $(\a,\b=e,\m)$ only one term 
contributes in the sum of (\ref{master}). The unitarity of the
mixing matrix $U$ then implies $(i=2, j=1)$,
$U_{\a 2}^* U_{\b 2} U_{\a 1} U_{\b 1}^* = -|U_{\a 2}|^2 |U_{\b 2}|^2$.
This yields for the transition probability
\begin{equation}\label{p2}
P(\n_\a \rightarrow \n_\b) = 
4 |U_{\a 2}|^2 |U_{\b 2}|^2 \sin^2{\left(\D m^2 {L\over 4E}\right)}\;,
\end{equation}
where $\D m^2 = m_2^2 - m_1^2$. The unitary $2\times 2$ mixing matrix can
be parametrized in the standard way in terms of 1 rotation angle and 
3 phases,
\begin{equation}\label{2f}
U = \left(\begin{array}{cc}
    e^{i\f_1}\cos{\Q_0}  &  e^{i(\f_2-\f_3)}\sin{\Q_0}  \\
    -e^{i(\f_1+\f_3)}\sin{\Q_0}  & e^{i\f_2}\cos{\Q_0} 
    \end{array}\right) \;.
\end{equation}
With $4 |U_{\a 2}|^2 |U_{\b 2}|^2 = \sin^2{2\Q_0}$ one then reads off from
eq.~(\ref{p2}) the standard formula for the transition probability in the
case of 2 flavours,
\begin{equation}\label{vacosc}
P(\n_\a \rightarrow \n_\b) = \sin^2{2\Q_0}\sin^2{{L\over L_{vac}}}\;,
\end{equation}
with the vacuum oscillation length
\beqa
L_{vac} &=& {4E\over \D m^2} = {2\over \D p} \NO\\
&\simeq & 
1.27\ \mbox{km} \left({E[\mbox{GeV}]\over \D m^2 [\mbox{eV}^2]}\right)\;.
\eeqa
For fast oscillations, i.e. large oscillation phase, averaging over the
energy resolution of the detector and the distance $L$ according to the 
uncertainty of the production point, yields the average transition probability,
\begin{equation}\label{avosc}
\overline{ P(\n_\a \rightarrow \n_\b)} = {1\over 2}\sin^2{2\Q_0}\;.
\end{equation}
In disappearance experiments one measures the survival probability
$P(\n_\a \rightarrow \n_\a)$ which is directly related to the transition
probability,
\beqa
P(\n_\a \rightarrow \n_\a) &=& 1 - P(\n_\a \rightarrow \n_\b) \NO\\
&=& 1 - 4 |U_{\a 2}|^2 (1 - |U_{\a 2}|^2)
\sin^2{\left(\D m^2 {L\over 4E}\right)}\;.
\eeqa
\item
Particularly interesting is the case of three hierarchical neutrinos,
$m_1^2 \ll m_2^2 \ll m_3^2$. Suppose that $\D m_{31}^2 L/(4E) = \co(1)$,
and correspondingly $\D m_{21}^2 L/(4E) \ll 1$. From eq.~(\ref{master})
one then obtains
\beqa
P(\n_\a \rightarrow \n_\b)  
&\simeq& - 4 \mbox{Re}\left(U_{\a 3}^* U_{\b 3} U_{\a 2} U_{\b 2}^*
                    +U_{\a 3}^* U_{\b 3} U_{\a 1} U_{\b 1}^*\right)
             \sin^2{\left(\D m^2_{31}{L\over 4E}\right)}\NO\\
&=& 4 |U_{\a 3}|^2 |U_{\b 3}|^2 \sin^2{\left(\D m^2_{31}{L\over 4E}\right)}\;.
\eeqa
Note, that this result is completely analogous to the two-flavour case; only
the large mass difference matters.

\item
The sensitivity of an oscillation experiment with respect to the neutrino
mass difference $\D m^2$ is determined by the neutrino energy $E$ and the
oscillation length $L$. Oscillations become visible for mass differences
above $\D m^2 L/(4E) \simeq 1$. In the appropriate units of energy and length
this condition reads
\begin{equation}
\D m^2 [\mbox{eV}^2] \simeq {E[\mbox{GeV}]\over L[\mbox{km}]}\;.
\end{equation}
Relevant examples are given in table~\ref{tab:osc}.

\begin{table}
\begin{center}
\begin{tabular}{l|ccc}
\hline \hline
& $L$[km] & $E$[GeV] & $\D m^2$ [eV$^2$] \\ \hline
accelerator (short baseline)  & 0.1 & 1         & 10  \\
reactor                       & 0.1 & $10^{-3}$ & $10^{-2}$ \\
accelerator (long baseline)   & $10^3$ & 10 & $10^{-2}$ \\
atmospheric                   & $10^4$ & 1 & $10^{-4}$ \\
solar                         & $10^8$ & $10^{-3}$ & $10^{-11}$ \\ \hline
\end{tabular}
\end{center}
\caption{{\it The approximate reach in $\D m^2$ of different oscillation
experiments.\label{tab:osc}}}
\end{table}

\item
In the past the sensitivity of neutrino oscillation experiments has been
mostly displayed in the ($\D m^2$, $\sin^2{2\Q}$) plane. Obviously, this
parametrization covers only half of the parameter space. More appropriate
are the variables ($\D m^2$, $\tan^2{\Q}$), which is particularly important
for oscillations in matter \cite{mur02}.  

\item
The measurement of magnitude and sign of $C\!P$ violation in neutrino oscillations
would be of fundamental importance. One $C\!P$ violating observable is the
asymmetry
\begin{equation}
\D_{e\m} = P(\n_e \rightarrow \n_\m) - P(\bar{\n}_e \rightarrow \bar{\n}_\m)\;.
\end{equation}
Using $C\!P\!T$ invariance and the unitarity of the mixing matrix one easily
verifies that in the case of three neutrino flavours the three possible
asymmetries are all equal,
\begin{equation}
\D_{e\m} = \D_{\m\t} = \D_{\t e}\;.
\end{equation}
From the master formula (\ref{master}) one reads off,
\begin{equation}
\D_{\a\b} = 4 \sum_{i>j} \mbox{Im}\left(
              U_{\a i}^* U_{\b i} U_{\a j} U_{\b j}^*\right)
             \sin^2{\left(\D m^2_{ij}{L\over 2E}\right)}\;.
\end{equation}
The product of mixing matrices is completely antisymmetric in the flavour
indices,
\begin{equation}\label{jarls}
\mbox{Im}\left(U_{\a i}^* U_{\b i} U_{\a j} U_{\b j}^*\right) =
\widetilde{\e}_{\a\b}\widetilde{\e}_{ij} J_l\;,
\end{equation}
where $\widetilde{\e}_{ab} = \sum_{c=1}^3 \e_{abc}$ and $J_l$ is the
leptonic Jarlskog parameter. For quarks one finds for the corresponding
quantity $J_q \sim 10^{-5}$. Due to the large neutrino mixings $J_l$
turns out to be much larger, which gives hope to observe $C\!P$ violating
effects in future superbeam experiments and at a neutrino factory \cite{cpv99}.

\end{itemize}

\subsection{Oscillations in matter}

In vacuum, neutrino oscillation probabilities are bounded by the mixing
angle, $P \leq \sin^2{2\Q_0}$. Hence, for small mixing angles transition
probabilites are small. In matter, a resonance enhancement of neutrino
oscillations can take place and transition probabilities can be maximal even 
for small vacuum mixing angles $-$ this is the Mikheyev-Smirnov-Wolfenstein
effect \cite{msw}.

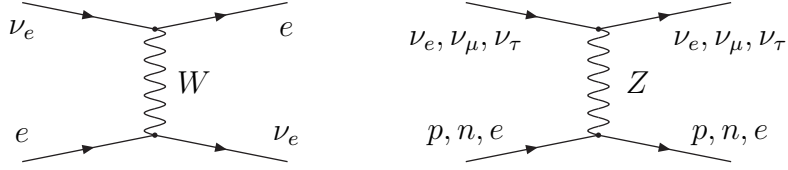
\begin{figure}
\begin{center}
\input{numat.tex}
\end{center}
\vspace{-0.3cm}
\caption{\label{fig:nupot}\it Neutrino interaction with matter via charged
and neutral currents.}
\end{figure}

The matter enhancement of oscillations is a coherent effect, due to elastic
forward scattering of neutrinos with negligible momentum transfer. The effect
can be described by the propagation of the neutrinos in an approximately
constant potential generated by the exchange of $W$-bosons 
(fig.~\ref{fig:nupot}),
\begin{equation}\label{numat}
L_{CC} = -2\sqrt{2} G_F \overline{e_L}\g^\m \n_{Le}
                         \overline{\n_L}\g_\m e_{L}\;.
\end{equation}
The corresponding $Z$-exchange does not distinguish between $\n_e$, $\n_\m$
and $\n_\t$, and therefore it has no influence on neutrino oscillations.

After a Fierz-transformation one obtains from eq.~(\ref{numat}),
\begin{equation}
L_{CC} = -2\sqrt{2} G_F \overline{\n_{Le}}\g^\m \n_{Le}
                         \overline{e_L}\g_\m e_L\;.
\end{equation}
In order to obtain the effective potential for the neutrino propagation in
matter, e.g. the interior of the sun, one has to evaluate the expectation
value of the electron current in the corresponding state, i.e. the quantity
$\langle \overline{e}\g^\m (1-\g_5)e\rangle$. Since no polarization vector
and no spatial direction are singled out, one has
\begin{equation}
\langle \overline{e}\g^\m \g_5 e\rangle = 0\;, \quad
\langle \overline{e}\g^i e\rangle = 0\;,
\end{equation}
\begin{equation}
\langle \overline{e}\g^0 e\rangle = 
\langle e^\dg e\rangle = 2 n_e \;,
\end{equation}
where $n_e$ is the electron number density and we have summed over electron
spins.

The final lagrangian describing the propagation of neutrinos now reads 
(cf.~(\ref{numass}), $m_\n=m_\n^*$),
\begin{equation}\label{majnum}
\cl = {1\over 2} \overline{\n_\a}\left(i\slpa \d_{\a\b} 
       - m_{\a\b}\right)\n_\b 
       - {G_F\over \sqrt{2}} n_e \overline{\n_e}\g^0\n_e\;.
\end{equation}
For simplicity we shall restrict ourselves in the following to the case of two
flavours, i.e. $\a,\b = e,\m$. We also consider the idealized case of
constant electron density. From (\ref{majnum}) one obtains the hamiltonian
\begin{equation}
H = H_0 + V\;,
\end{equation} 
with
\begin{equation}
H_0 = \g^0\vec{\g}{1 \over i} \vec{\partial} + \g^0 m\;, \quad
V = \left(\begin{array}{cc} \sqrt{2}G_F n_e & 0 \\ 0 & 0 
\end{array}\right)\;.
\end{equation}
Note, that the matter induced potential has the same Lorentz structure as
a static Coulomb potential. Hence, matter effects have opposite sign
for neutrinos and anti-neutrinos.

The probability for flavour oscillations can now be calculated completely
analogous to the case of vacuum oscillations. We again have to find a
stationary state,
\begin{equation}
H \n_E^{(\a)}(z) = E \n_E^{(\a)}(z) \;,
\end{equation}
which satisfies the boundary condition,
\begin{equation}
\n_E^{(\a)}(0) = \n_\a \;. \NO
\end{equation}
The free hamiltonian is known in the mass eigenstate basis (cf.~(\ref{massham})) 
where we again neglect the mixing left- and right-handed states. 
In the flavour basis one then has
\begin{equation}
H = U(\Q_0) H_0 U^\dg(\Q_0) + V \;.
\end{equation}
With
\begin{equation}
U(\Q_0) = \left(\begin{array}{cc} \cos{\Q_0} & \sin{\Q_0} \\
           -\sin{\Q_0} &  \cos{\Q_0} \end{array}\right)\;,
\end{equation}
the hamiltonian reads explicitely
\begin{equation}\label{hmat}
H \simeq \left(\begin{array}{cc} 
\hat{p}+{m_1^2+m_2^2\over 4E}-{\D m^2\over 4E}\cos{2\Q_0}+\sqrt{2}G_F n_e &
{\D m^2\over 4E}\sin{2\Q_0} \\ {\D m^2\over 4E}\sin{2\Q_0} &
\hat{p}+{m_1^2+m_2^2\over 4E}+{\D m^2\over 4E}\cos{2\Q_0} \end{array}\right)\;.
\end{equation}
Here $\D m^2 = m_1^2 - m_2^2$, and we have assumed that the relevant eigenvalues 
of $\hat{p}$ are $E + \co\left(m_i^2/E\right)$.
It is now straightforward to diagonalize $H$, requiring
\begin{equation}
U^\dg(\Q) H U(\Q)=\left(\begin{array}{cc} E & 0 \\ 0 & E \end{array}\right)\;.
\end{equation}
From (\ref{hmat}) one obtains the rotation angle
\begin{equation}\label{matang}
\tan{2\Q} = {{\D m^2\over 2E}\sin{2\Q_0}\over {\D m^2\over 2E}\cos{2\Q_0}
             - \sqrt{2}G_F n_e}\;,
\end{equation}
and the equation for the momentum eigenvalues
\begin{equation}
\left(\begin{array}{cc} 
p_1+{m_1^2+m_2^2\over 4E}+\sqrt{2}G_F n_e - {\D\over 2} & 0 \\
0 & p_2+{m_1^2+m_2^2\over 4E}+\sqrt{2}G_F n_e + {\D\over 2}
\end{array}\right) 
= \left(\begin{array}{cc} E & 0 \\ 0 & E \end{array}\right)\;,
\end{equation}
where
\begin{equation}
\D = {\D m^2\over 2E}\cos{2\Q_0} - \sqrt{2}G_F n_e\;. 
\end{equation}
This yields the momentum eigenvalues
\beqa
p_1 + p_2 &=& 2E - {m_1^2 + m_2^2\over 2E} - \sqrt{2}G_F n_e \;,
\label{sum}\\
p_1 - p_2 &=& \D p\ = \left(
 \left({\D m^2\over 2E}\cos{2\Q_0} - \sqrt{2}G_F n_e\right)^2 + 
 \left({\D m^2\over 2E}\sin{2\Q_0}\right)^2\right)^{1/2}\;.\label{dif}
\eeqa
\begin{figure}
\begin{center}
\input{msw.tex}
\end{center}
\vspace{-1cm}
\caption{\label{fig:msw}\it Change of neutrino momentum and flavour in the sun.}
\end{figure}
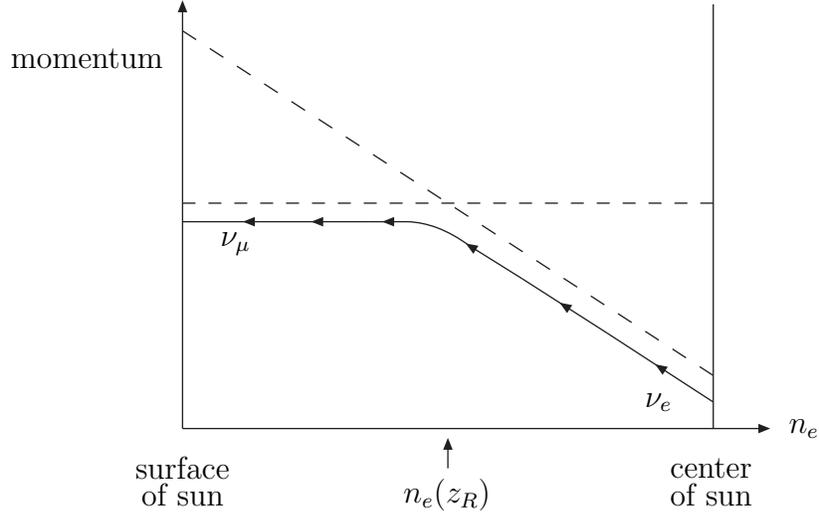

The probability for flavour oscillations is again given by eq.~(\ref{vacosc}), 
with $\Q_0$ replaced by $\Q$ and $L_{vac}$ replaced by $L_{mat} = 2/\D p$,
\beqa
P(\n_\a \rightarrow \n_\b) &=& |\n_\b^\dg \n_E^{(\a)}(z)|^2 \NO\\
&\simeq& \sin^2{2\Q} \sin^2{{L\over L_{mat}}}\;.
\eeqa
The expression (\ref{matang}) for the rotation angle has the typical
resonance form, and maximal mixing, i.e. $\Q = 45^o$, is reached for
\begin{equation}\label{res}
\sqrt{2}G_F n_e = {\D m^2\over 2E}\cos{2\Q_0} \;.
\end{equation}
This is the MSW resonance condition. It requires in particular that
${\D m^2\over 2E}\cos{2\Q_0}$ is positive. A negative sign would allow
resonant oscillations for antineutrinos.

At the center of the sun the electron density is 
$n_e \sim 10^{26}/\mbox{cm}^3$. With $E \sim 1$~MeV, the MSW resonance
condition can be fullfilled on the way out to the surface of the sun for
mass differences $\D m^2 \sim 10^{-5}$~eV$^2$. In the `adiabatic 
approximation', where the oscillations are fast compared to the 
variation of the electron density, eqs.~(\ref{sum}), (\ref{dif}) and
(\ref{res}) can be used at each distance $z$ from the center. The physical
picture is particularly simple in the case of small mixing angle $\Q_0$,
where the off-diagonal terms in the hamiltonian (\ref{hmat}) are small.
The momentum of the produced electron neutrino increases, and near $z_R$ the
transition $\n_e \rightarrow \n_\m$ takes place after which the momentum 
stays constant (fig.~\ref{fig:msw}). A more accurate calculation of the
oscillation probability can be carried out similar to the treatment of the
Landau-Zener effect in atomic physics \cite{akh99,ggn02}.

\subsection{Comparison with experiment}

We are now ready to interpret the deficit in the solar and atmospheric
neutrino fluxes in terms of neutrino oscillations. In the following only
a brief account of the main results will be given. More detailed
discussions of the fusion processes in the sun, solar neutrino experiments,
the atmospheric neutrino anomaly, reactor experiments and global fits
can be found in \cite{bgg99}-\cite{ggn02}.

\subsubsection{Solar neutrinos}

\begin{figure}
\centerline{\epsfig{file=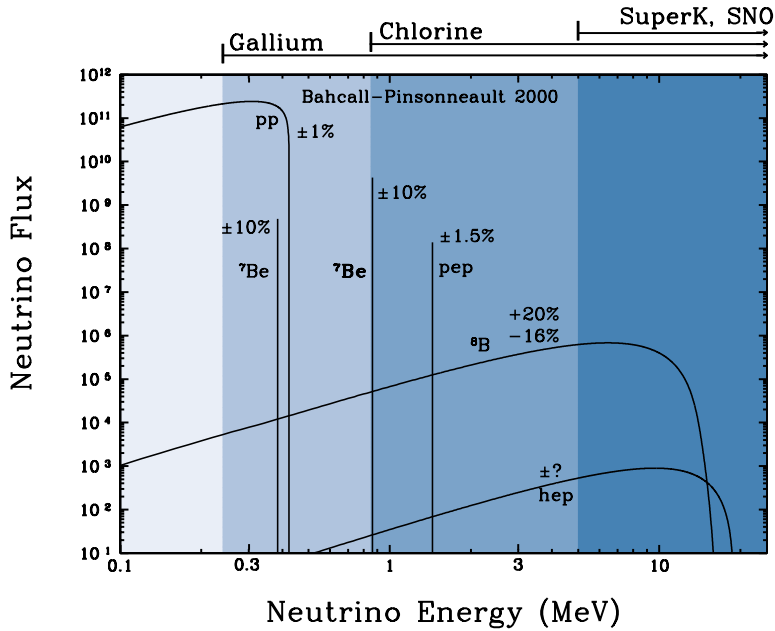,width=9cm,angle=270}}
\caption{\it Solar neutrino spectrum and energy thresholds of solar
neutrino experiments \protect\cite{bahhome}.\label{fig:snuflux}}
\end{figure}

The solar energy is generated by thermonuclear reactions, in particular
the $pp$ cycle and the CNO cycle, which also produce electron neutrinos.
The largest neutrino flux, 
$\Phi_{\n_e} = 6.0\times 10^{10}$/[cm$^2$s] is due to the reaction,
\begin{equation}\label{dom}
p + p \rightarrow d + e^+ + \n_e \;,
\end{equation}
with a maximum neutrino energy of 0.42~MeV. Also important are the processes
$e^- +\ ^7Be \rightarrow \n_e +\ ^7Li$ with 
$\Phi_{\n_e} = 4.9\times 10^9$/[cm$^2$s], $E_\n = 0.86$~MeV 
and $^8B \rightarrow \ ^8Be + e^+ + \n_e$, with 
$\Phi_{\n_e} = 5.0\times 10^6$/[cm$^2$s], $E_\n \leq 0.86$~MeV. 
The complete energy spectrum of solar neutrinos is summarized in 
fig.~\ref{fig:snuflux}.

\begin{figure}
\centerline{\epsfig{file=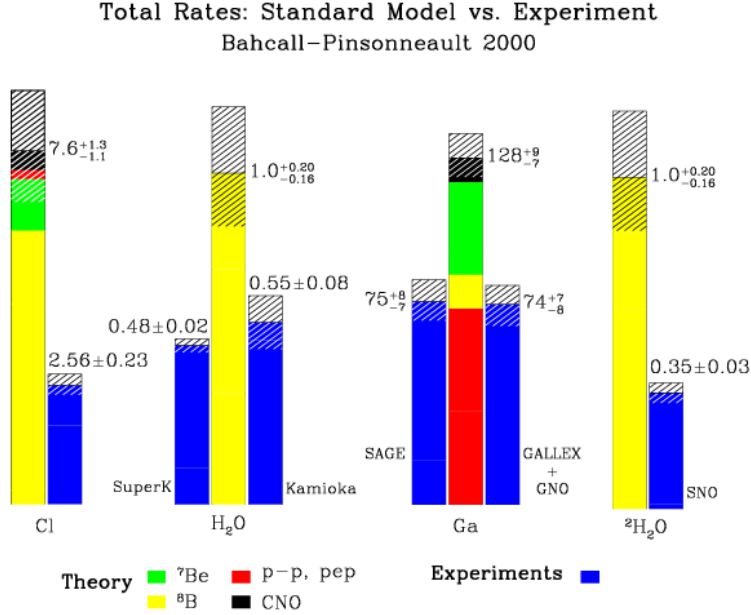,width=9cm,angle=270}}
\caption{\it Theoretical flux predictions compared with experimental results; 
for the Cl and Ga experiments the units are SNU, for Kamiokande and 
Superkamiokande the ratio of measured and predicted fluxes are shown
\protect\cite{bahhome}.\label{fig:snuexp}}
\end{figure}

The deficit in the solar neutrino flux was discovered in radiochemical
experiments. The first one was the Homestake experiment of Davis et al.
based on the reaction
\begin{equation}
\n_e +\ ^{37}Cl \rightarrow\ ^{37}Ar + e^-\;,
\end{equation}
with an energy threshold of 0.814 MeV. Only 1/3 of the predicted flux was
observed. 

The GALLEX, SAGE and GNO experiments detect solar neutrinos via the reaction
\begin{equation}
\n_e +\ ^{71}Ga \rightarrow\ ^{71}Ge + e^-\;.
\end{equation}
Since the energy threshold is only 0.234 MeV, these
experiments could see for the first time neutrinos from the dominant
process (\ref{dom}). About 60\% of the predicted flux is observed in
this reaction (fig.~\ref{fig:snuexp}).

The Kamiokande and Super-Kamiokande experiments are water Cherenkov
detectors. Here the reaction,
\begin{equation}
\n_x + e^- \rightarrow \n_x + e^- \;,
\end{equation}
is used to detect solar neutrinos. To suppress background only high-energy
neutrinos from the $^8B$ decay, with $E>7$~GeV, are detected. As a consequence 
the recoil electrons are peaked in the direction of the incoming neutrino.
The angular distribution of the detected electrons shows indeed a peak
opposite to the direction to the sun. This clearly demonstrates the solar
origin of detected neutrinos!
\begin{figure}
\begin{center}
\input{el.tex}
\end{center}
\vspace{-1cm}
\caption{\it Elastic neutrino electron scattering via
charged and neutral currrents.\label{fig:el}}
\end{figure}
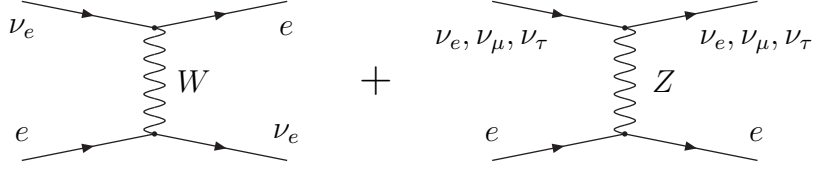

An important step towards the final resolution of the solar neutrino problem
was made by the recent results of the Sudbury Neutrino Observatory (SNO).
Like in Super-Kamiokande, high-energy $^8B$ neutrinos are observed, this
time in a 1000 t detector of heavy water $D_2 O$. This allows the observation
of the charged current (CC) reaction,
\begin{equation}\label{ccreac}
\n_e + d \rightarrow p + p + e^- \;,
\end{equation}
as well as the elastic scattering (ES),
\begin{equation}\label{esreac}
\n_a + e^- \rightarrow \n_a + e^- \;,
\end{equation}
\begin{figure}[h]
\centerline{\epsfig{file=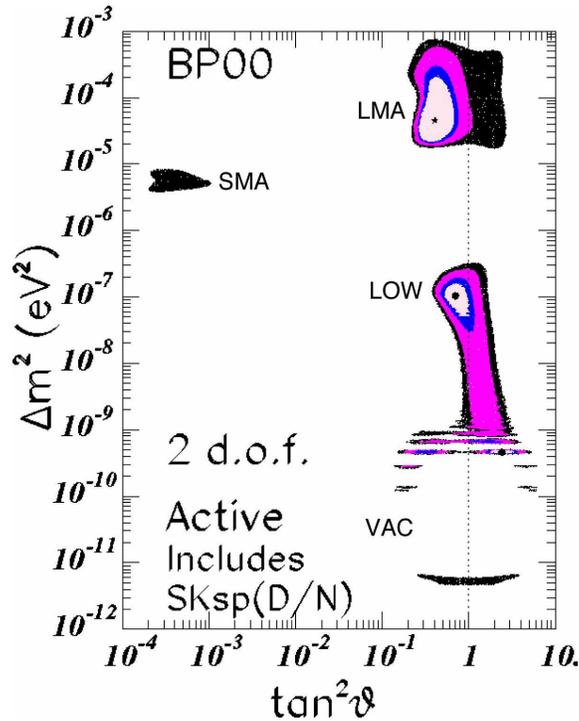,width=7.5cm}}
\caption{\it Two-flavour ($\n_e$ to $\n_a$) fit of solar neutrino data, with 
confidence levels 90\%, 95\%, 99\%, 99.7\% \protect\cite{bgp01}.
\label{fig:twofla}}
\end{figure}
\begin{figure}[h]
\centerline{\epsfig{file=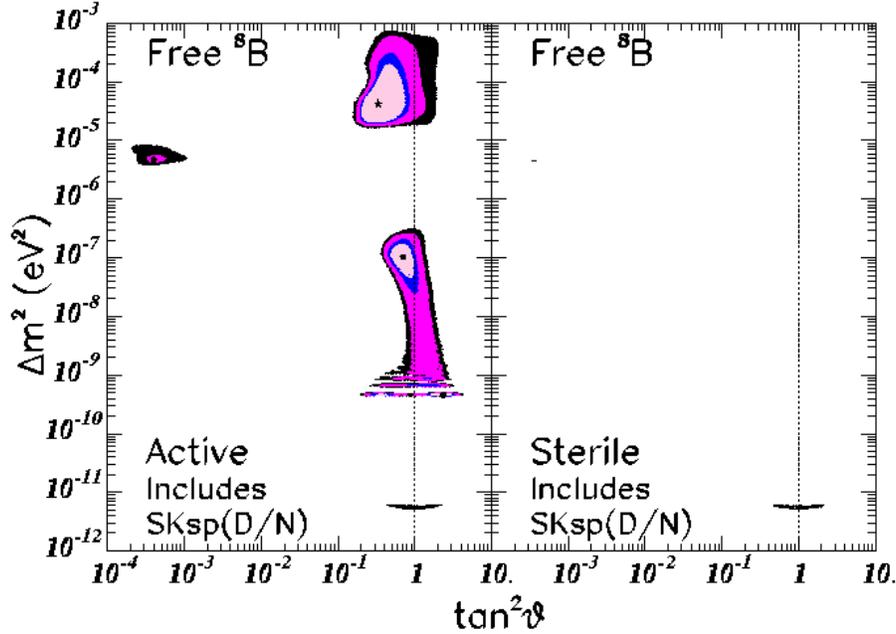,width=12cm}}
\vspace{-0.5cm}
\caption{\it Comparison of two-flavour $\n_e \rightarrow \n_a$ and  
$\n_e \rightarrow \n_s$ fits, leaving the $^8B$ flux free \protect\cite{bgp01}.
\label{fig:acste}}
\end{figure}
which involves charged and neutral currents (fig.~\ref{fig:el}).
From the CC reaction (\ref{ccreac}) one can, for the first time (!), directly 
determine the $\n_e$ flux \cite{sno01},
\begin{equation}\label{fluxCC}
\Phi^{CC}(\n_e) = 
1.75\pm 0.07 (stat)^{+0.12}_{-0.11}(syst)\pm 0.05 (theo)\times 10^6 
\mbox{cm}^{-2}\mbox{s}^{-1}\;.
\end{equation}
Here statistical, systematic and theoretical errors are listed separately.
Under the assumption that the entire flux consists of electron neutrinos,
one obtains from the ES process \cite{sno01},
\begin{equation}\label{fluxES}
\Phi^{ES}(\n_e) = 
2.39\pm 0.34 (stat)^{+0.16}_{-0.14}(syst)\times 10^6 
\mbox{cm}^{-2}\mbox{s}^{-1}\;.
\end{equation}
Comparing $\Phi^{CC}$ and $\Phi^{ES}$ one concludes that with a significance
of 1.6 $\s$ there is evidence for a non-$\n_e$ component in the solar neutrino
flux. Combining this with the more precise measurement of $\Phi^{ES}$ by
Superkamiokande the significance increases to 3.3 $\s$.

Alternatively, one may assume that electron-neutrinos have partially been
converted into muon- and tau-neutrinos. In this case, the flux extracted
from the elastic scattering (\ref{esreac}) should coincide with the
flux of $^8B$ neutrinos. One finds \cite{sno01},
\begin{equation}\label{fluxB}
\Phi(\n_a) = 5.44\pm 0.99 \times 10^6 \mbox{cm}^{-2}\mbox{s}^{-1}\;,
\end{equation}
in remarkable agreement with the prediction of the standard solar model! 

Global fits of all solar neutrino data have been performed under different
assumptions \cite{bgp01,flx01}. Fig.~\ref{fig:twofla} shows a two-flavour
fit for the oscillation of $\n_e$ into an active neutrino $\n_a$ using
the solar flux PB2000 (fig.~\ref{fig:snuflux}). Four regions in
parameter can describe the data: LMA (Large Mixing Angle MSW solution),
SMA (Small Mixing Angle MSW solution), LOW (MSW solution with low $\D m^2$)
and VAC (Vacuum oscillation solution). The LMA solution gives the best fit.

In fig.~\ref{fig:acste} two-flavour fits for oscillations into active and
sterile neutrinos are compared. Clearly, the oscillation 
$\n_e \rightarrow \n_s$ is essentially excluded. For the oscillation into
a superposition of active and sterile neutrino the best fit is obtained
for vanishing sterile component.

\subsubsection{Atmospheric neutrinos}

Cosmic rays produce in the earth's atmosphere pions and kaons which decay
into charged leptons and neutrinos,
\beqa
\p^{\pm}, K^{\pm} &\rightarrow& \m^{\pm} + \n_\m (\overline{\n}_\m)\;, \NO\\
&&\m^{\pm}\rightarrow e^{\pm}+\n_e(\overline{\n}_e)+\overline{\n}_\m (\n_\m)\;.
\eeqa
These neutrinos can be detected by the usual neutrino-nucleon charged current
reactions. 
Naive counting suggests that the corresponding flux of `atmospheric neutrinos'
contains twice as many muon-neutrinos than electron-neutrinos. More
accurately, one compares the measured ratio of $\n_\m$'s and $\n_e$'s, 
$N_\m/N_e$, with the theoretical expectation of this quantity.
For the double ratio,
\begin{equation}
R = {\left(N_\m/N_e\right)_{data}\over\left(N_\m/N_e\right)_{MC}}\;,
\end{equation}
the Super-Kamiokande collaboration finds \cite{ska01},
\begin{equation}
R = 0.638 \pm 0.017 (stat) \pm 0.050 (syst) \;,
\end{equation}
where the sample has been restricted to sub-GeV ($E< \sim 1$~GeV) charged 
leptons. For the sample of multi-GeV ($E> \sim 1$~GeV) leptons a similar 
number is found.
\begin{figure}
\centerline{\epsfig{file=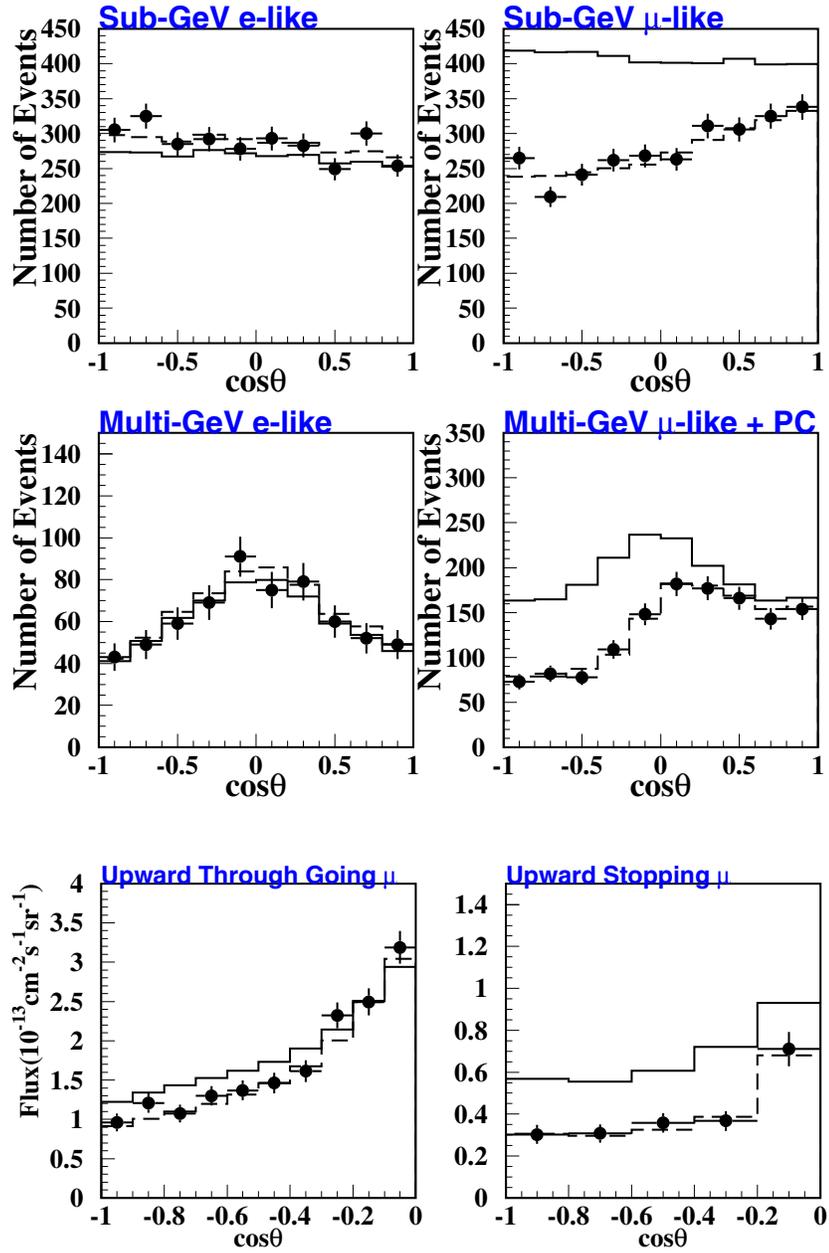,width=12cm}}
\caption{\it Zenith angle distribution of Super-Kamiokande 1289 day
sub-GeV and multi-GeV samples. Solid and dashed lines correspond to
MC with no oscillation and MC with best fit oscillation parameters,
respectively \protect\cite{ska01}.\label{fig:zenith}}
\end{figure}

It appears natural to interpret the observed $\n_\m$ deficit again in terms of
neutrino oscillations. This hypothesis is further strengthened by the
analysis of the zenith angle dependence shown in fig.~\ref{fig:zenith}.
No deficit is seen for electron-neutrinos. On the contrary, for muon-neutrinos
the deficit is the larger, the larger the zenith angle, i.e. the larger the
distance between detector and production point in the earth's atmosphere.
For upward going neutrinos the oscillation length is about 13.000 km.

For neutrinos with energies $\co$(GeV) one has a sensitivity for mass
splittings down to $\D m^2 \sim 10^{-4}$~eV$^2$. A two-flavour analysis
$\n_\m \rightarrow \n_a$ yields for the allowed parameter range at
90\% CL,
\begin{equation}
\sin^2{2\Q_0} > 0.88\;, \quad 1.6\times 10^{-3} < \D m^2 < 4\times 10^{-3}\;.
\end{equation}
For oscillations into  sterile neutrinos, $\n_\m \rightarrow \n_s$, earth  
matter effects become important. A detailed analysis shows that this case
is disfavoured.

Important information on neutrino mixing is also obtained from reactor experiments.
Failure to observe $\overline{\n}_e \rightarrow \overline{\n}_x$ oscillations
gives the bound \cite{rea01}, for $\D m^2_{31} \simeq 3\times 10^{-3}$~eV$^2$,
\begin{equation}
\sin^2{2\Q_0} < 0.12\;.
\end{equation}
For three active neutrinos, this suggests a small mixing between the first and
the third generation.

\subsubsection{Future prospects}

In the coming years we can look forward to important results from several
experiments. Mini-BooNE at Fermilab will verify or falsify the LSND
signal for oscillations. The long-baseline experiments K2K and MINOS
will be able to verify $\n_\m - \n_\a$ oscillations and to determine the 
mass difference $\D m^2_{23}$ more precisely. Correspondingly, OPERA and
ICARUS at Gran Sasso, using the muon-neutrino beam from CERN, will look for
$\t$ appearance, verifying unequivocally $\n_\m - \n_\t$ oscillations.
The first oscillation dip could be studied by MONOLITH. The $^7Be$ line
of solar neutrinos will be studied by Borexino.

\begin{figure}
\centerline{\epsfig{file=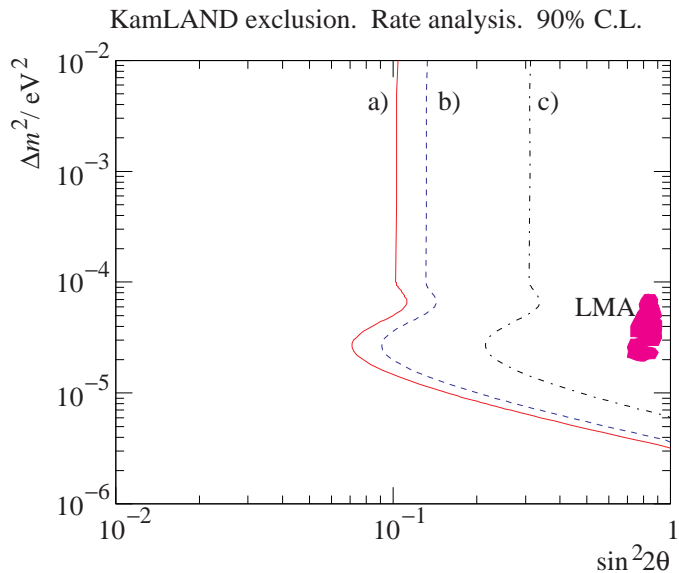,width=9cm}}
\caption{\it Sensitivity of the KamLAND neutrino oscillation experiment.
The curves represent 90\% CL limits which can be reached in 3 years of
running. The cases a) - c) correspond to different assumptions about the
background \protect\cite{kam99}.\label{fig:kaml}}
\end{figure}

KamLAND is the first terrestrial experiment sensitive to the solar neutrino
signal (fig.~\ref{fig:kaml}). Using reactor neutrinos over a baseline of 
about 175 km it could
verify the LMA solution of the solar neutrino problem. This would be of
crucial importance for the prospects of observing CP violation in neutrino  
oscillations. This may then be possible in neutrino superbeam experiments
(JHF in Japan, SPL at CERN) or, eventually, at a neutrino factory.

The result of neutrino oscillation experiments will be the determination of
neutrino mass differences and the leptonic mixing matrix. At present, a
three-flavour analysis assuming the LMA solution leads to the following result 
\cite{ft01},
\begin{equation}
U = \left(\begin{array}{ccc} 0.74 - 0.90 & 0.45 - 0.65 & <0.16\\
0.22 - 0.61 & 0.46 - 0.77 & 0.57 - 0.71 \\
0.14 - 0.55 & 0.36 - 0.68 & 0.71 - 0.82 \end{array}\right)\;.
\end{equation}
The emerging structure is very remarkable and completely different from the
CKM matrix of quark mixing. Whereas $V_{CKM}$ is strongly hierarchical, all
elements of the leptonic mixing matrix $U$ are of the same order, except
$U_{e3}$, for which only an upper bound exists. This has important implications
for the structure of fermion masses in unified theories.

\section{Neutrino Masses in GUTs}

Neutrino masses and their relation to quark and charged lepton masses play
an important role in grand unified theories. According to the seesaw mechanism
the smallness of the light neutrino masses is explained by the largeness of
the heavy Majorana neutrino masses, which extend up to the mass scale of
unification. Hence, the present experimental evidence for neutrino masses and 
mixings probes for the first time the physics of unification. Reviews and
extensive references are given in ref.~\cite{rev02}, a special approach is
described in ref.~\cite{fnt02}. 

\subsection{Elements of grand unified theories}

Let us first consider the basic ingredients of GUTs, starting from the
symmetries of the standard model where neutrinos are massless. 
As a consequence, four `charges' are classically
conserved, three lepton numbers and baryon number,
\begin{equation}
L_e\;,\quad L_\m\;, \quad L_\t\;, \quad B\;.
\end{equation} 
\begin{figure}[h]
\begin{center}
\input{ano.tex}
\end{center}
\vspace{-0.5cm}
\caption{\it ABJ-anomaly of the $B-L$ current.\label{fig:ano}}
\end{figure}
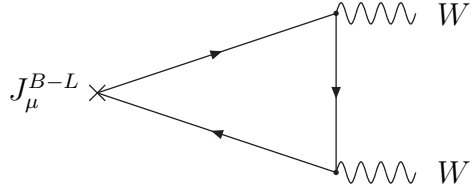
The Adler-Bell-Jackiw anomaly (fig.~\ref{fig:ano}), a quantum effect, 
reduces the four conserved charges to three \cite{tho76}, which may be 
chosen as,
\begin{equation}
L_e-L_\m\;, \quad L_\m-L_\t\;, \quad B-L\;,
\end{equation}
where $L=L_e + L_\m + L_\t$ is the total lepton number. One may wonder
whether these charges correspond to fundamental global symmetries, or whether
they are just accidental symmetries of the low-energy effective theory,
similar to weak isospin which is an approximate symmetry of weak interactions
at energies much below the W-boson mass.

\begin{figure}
\begin{center}
\input{uni.tex}
\end{center}
\vspace{-0.5cm}
\caption{\it Unification of gauge couplings in the supersymmetric standard
model.\label{fig:uni}}
\end{figure}
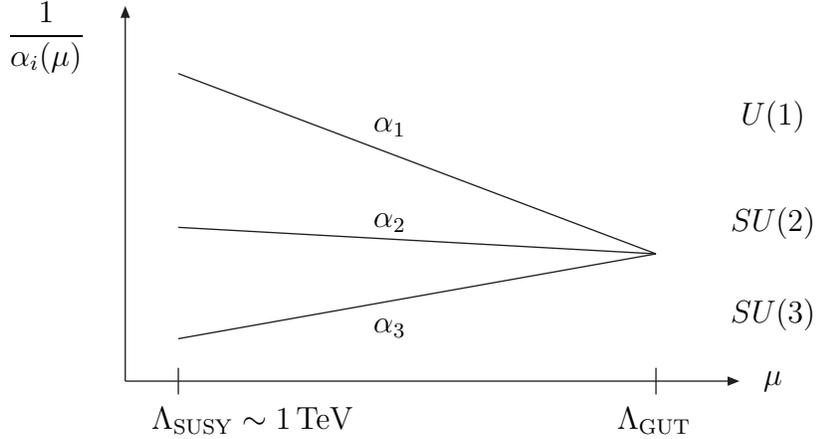

Neutrino masses and mixings break the first two of these symmetries,
$L_e-L_\m$ and $L_\m-L_\t$, but presently we do not know whether $B-L$
is a fundamental symmetry or not. $B-L$ is conserved if neutrinos have
only Dirac masses and the Majorana masses of the right-handed neutrinos
vanish,
\begin{equation}
m_D = h_\n \langle H_1\rangle\;, \quad M = 0\;.
\end{equation}
Note, that this requires tiny Yukawa couplings, e.g. $h_{\n22} \sim 10^{-14}$
for $m_{\n2} \sim 5\times 10^{-3}$~eV. This possibility can presently not be
excluded although it might appear unnatural.
Alternatively, $B-L$ may be broken at some intermediate mass scale or at the
GUT scale $\L \sim 10^{16}$~GeV, where the gauge couplings unify in the
supersymmetric standard model (fig.~\ref{fig:uni}). This is indeed
suggested by the smallness of the observed neutrino masses in connection
with the seesaw mechanism. With $h_{\n33} = \co(1)$ one obtains for the mass
of the heaviest Majorana neutrino from eq.~(\ref{wettss}),
\begin{equation}
M_3 \sim {\langle H_1\rangle^2 \over m_{\n_3}} \sim 10^{15}\ \mbox{GeV}
\sim \L_{GUT}\;.
\end{equation}

The unification of gauge couplings suggests that the standard model
gauge group is part of a larger simple group,
\begin{equation}
G_{SM} = U(1)\times SU(2)\times SU(3) \subset SU(5) \subset \ldots\;.
\end{equation}
The simplest GUT is based on the gauge group SU(5) \cite{gg74}. Here
quarks and leptons are grouped into the multiplets,
\begin{equation}
{\bf 10} =(q_L,u_R^c,e_R^c)\;, \quad 
{\bf 5^*} = (d_R^c,l_L)\;, \quad {\bf 1}= \n_R\;.
\end{equation}
Note that, unlike gauge fields, quarks and leptons are not unified in a 
single irreducible representation. In particular, the right-handed neutrinos 
are gauge singlets and can therefore have Majorana masses not generated by
spontaneous symmetry breaking. In addition one has three Yukawa interactions, 
which couple the fermions to the Higgs fields $H_1({\bf 5})$ and 
$H_2({\bf 5^*})$, 
\begin{equation} 
{\cal L} = h_{uij} {\bf 10}_i {\bf 10}_j H_1({\bf 5})
          +h_{dij} {\bf 5^*}_i {\bf 10}_j H_2({\bf 5^*}) 
          +h_{\n ij} {\bf 5^*}_i {\bf 1}_j H_1({\bf 5})
          + M_{ij} {\bf 1}_i {\bf 1}_j \;.  
\end{equation}
The mass matrices of up-quarks, down-quarks, charged leptons and the Dirac
neutrino mass matrix are given by
$m_u = h_u v_1$, $m_d = h_d v_2 = m_e$ and $m_D = h_\n v_1$, 
respectively, with $v_1 = \langle H_1\rangle$ and 
$v_2 = \langle H_2\rangle$. The SU(5) mass
relation $m_d = m_e$ is successful for the third generation but requires
substantial corrections for the second and first generations.
The Majorana masses $M$ are independent of the
Higgs mechanism and can therefore be much larger than the electroweak scale
$v$.

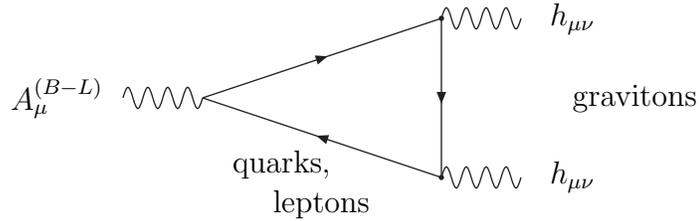
\begin{figure}[h]
\begin{center}
\input{grav.tex}
\end{center}
\vspace{-0.5cm}
\caption{\it Mixed $B-L$ $-$ gravitational anomaly.\label{fig:grav}}
\end{figure}

Once right-handed neutrinos are introduced, the $B-L$ charges of each 
generation add up to zero. Hence, there are no mixed
$B-L$ $-$ gravitational anomalies (fig.~\ref{fig:grav}), and the U(1)$_{B-L}$ 
symmetry can be embedded together with SU(5) into a larger GUT group.

In this way one arrives at the gauge group SO(10) \cite{gfm75}. All quarks and
leptons of one generation are now contained in a single multiplet,
\begin{equation}
{\bf 16}=(q_L,u_R^c,e_R^c,d_R^c,l_L,\n_R)\;.
\end{equation}
Quark and lepton mass matrices are obtained from the couplings of the 
fermion  multiplets ${\bf 16}_i$ to the Higgs 
multiplets $H_1({\bf 10})$, $H_2({\bf 10})$ and $\Phi({\bf 126})$,
\begin{equation} 
{\cal L} = h_{uij} {\bf 16}_i {\bf 16}_j H_1({\bf 10})
          +h_{dij} {\bf 16}_i {\bf 16}_j H_2({\bf 10})
          +h_{Nij} {\bf 16}_i {\bf 16}_j \Phi({\bf 126})\;.
\end{equation}
Here we have assumed that the two Higgs doublets of the standard model
are contained in the two {\bf 10}-plets $H_1$ and $H_2$, respectively.
This yields the quark mass matrices $m_u = h_u v_1$, $m_d = h_d v_2$, 
with $v_1 = \langle H_1\rangle$ and $v_2 = \langle H_2\rangle$, and the 
lepton mass matrices
\begin{equation}\label{muni}
m_D = m_u  \;, \quad m_e = m_d \;.
\end{equation}
Contrary to SU(5) GUTs, the Dirac neutrino and the up-quark mass matrices
are now related. Note, that all matrices are symmetric. The Majorana mass 
matrix $M = h_N \VEV \Phi$, which is also generated by spontaneous symmetry,
is a priori independent of $m_u$ and $m_d$. 

In the literature also GUT groups larger than SO(10) have been studied. 
Particularly interesting
is the sequence of exceptional groups which terminates at rank 8,
\begin{equation}
E_4 = SU(5) \subset E_5 = SO(10) \subset E_6 \subset E_7 \subset E_8\;.
\end{equation}
The last two groups can also unify different generations and thereby restrict
the Yukawa matrices. They arise naturally in higher dimensional supergravity
theories and in string theories.

What are the GUT predictions for neutrino masses and mixings? As emphasized 
at the end of section~4.3.3, the emerging structure of the leptonic mixing
matrix appears to be remarkably simple,
\begin{equation}\label{umix}
U = \left(\begin{array}{ccc}
    \ast  & \ast  & \diamond \\
    \ast  & \ast  & \ast \\
    \ast  & \ast  & \ast 
    \end{array}\right) \;,
\end{equation}
where the `$\ast$' denotes matrix elements whose value is consistent with the 
range $0.5 \ldots 0.8$, whereas for the matrix element 
`$\diamond$' only an upper bound exits, $|U_{e3}| < 0.16$. 

There are several interesting `Ans\"atze' to explain this pattern, such as
`bi-maximal' mixing ($\Q_{12} = \Q_{23} = 45^o$, $\Q_{13} = 0$),
\begin{equation}
U = \left(\begin{array}{ccc} {1\over \sqrt{2}} & -{1\over \sqrt{2}} & 0\\
{1\over 2} & {1\over 2} & -{1\over \sqrt{2}} \\
{1\over 2} & {1\over 2} & {1\over \sqrt{2}} \end{array}\right)\; ,
\end{equation}
or `democratic mixing' where in the weak eigenstate basis all elements
are equal to 1, which leads to the mixing matrix,
\begin{equation}
U = \left(\begin{array}{ccc} {1\over \sqrt{2}} & -{1\over \sqrt{2}} & 0\\
{1\over \sqrt{6}} & {1\over \sqrt{6}} & -{2\over \sqrt{6}}\\
{1\over \sqrt{3}} & {1\over \sqrt{3}} & {1\over \sqrt{3}}\end{array}\right)\;.
\end{equation}
Further, one can also construct `tri-bimaximal' mixing \cite{hps02}. All these 
patterns are interesting. In general, however, they lack an underlying symmetry.
In the following we shall restrict our discussion to constraints on neutrino 
masses which arise for the simplest GUT groups, SU(5) and SO(10).  

An important consequence of neutrino mixing are flavour changing processes
and electric dipole moments of charged leptons \cite{fla02}. In particular
supersymmetric theories predict effects large enough to be discovered in the
near future, even before the start of LHC.

\subsection{Models with SU(5)}

An attractive framework to explain the observed mass hierarchies of quarks
and charged leptons is the Froggatt-Nielsen mechanism \cite{fn79} based
on a spontaneously broken U(1)$_F$ generation symmetry. 
The Yukawa couplings are assumed to
arise from non-renormalizable interactions after a gauge singlet field $\F$ 
acquires a vacuum expectation value,
\begin{equation}
h_{ij} = g_{ij} \left({\VEV\F\over \L}\right)^{Q_i + Q_j}\;.
\end{equation}
Here $g_{ij}$ are couplings ${\cal O}(1)$, and $Q_i$ are the U(1)$_F$ 
charges of the
various fermions, with $Q_{\F}=-1$. The interaction scale $\L$ is
usually chosen to be very large, $\L > \L_{GUT}$. 
\begin{table}[b]

\begin{center}
\begin{tabular}{c|ccccccccc}\hline \hline
$\j_i$  & $ \bf 10_3 $ & $ \bf 10_2 $ & $ \bf 10_1 $ & $ \bf 5^*_3 $ & 
$ \bf 5^*_2 $ & $\bf 5^*_1 $ & $ \bf 1_3 $ & $ \bf 1_2 $ & $ \bf 1_1 $ 
\\\hline
$Q_i$  & 0 & 1 & 2 & $a$ & $a$ & $a+1$ & b & $c$ & $d$ \\ \hline\hline
\end{tabular}
\end{center}
\caption{\label{tab:lop}\it Lopsided U(1)$_F$ charges of SU(5) multiplets.}
\end{table}

The symmetry group SU(5)$\times$U(1)$_F$ has been considered by a number of 
authors. Particularly interesting is the case with a `lopsided' family 
structure where the chiral U(1)$_F$ charges are different for the 
$\bf 5^*$-plets and the $\bf 10$-plets of the same 
family \cite{sy98}-\cite{by99}. Note, that such lopsided charge assignments 
are not
consistent with the embedding into a higher-dimensional gauge group, like
SO(10)$\times$U(1)$_F$ or E$_6\times$U(1)$_F$.
An example of phenomenologically allowed lopsided charges $Q_i$ is given in 
table~\ref{tab:lop}.

The charge assignements determine the structure of the Yukawa matrices,
e.g.,
\begin{equation}\label{yuk}
h_e, h_\n
  \ \sim\  \left(\begin{array}{ccc}
    \e^3 & \e^2 & \e^2 \\
    \e^2 &\;  \e \;   & \e   \\
    \e   & 1    & 1
    \end{array}\right) \;,
\end{equation}
where the parameter $\e = \VEV\F/\L$ controls the flavour mixing, and
coefficients ${\cal O}(1)$ are unknown.
The corresponding mass hierarchies for up-quarks, down-quarks and charged
leptons are,
\beqa
\qquad\quad 
m_t : m_c : m_u  &\simeq & 1 : \e^2 : \e^4\;, \\
m_b : m_s : m_d  &=& m_\t : m_\m : m_e \simeq 1 : \e : \e^3\;.
\eeqa
The differences between the observed down-quark mass hierarchy and the
charged lepton mass hierarchy can be accounted for by introducing 
additional Higgs fields \cite{gj79}. From a fit to the running quark and
lepton masses at the GUT scale the flavour mixing parameter is determined  
as $\e \simeq 0.06$.

The light neutrino mass matrix is obtained from the seesaw formula,
\begin{equation}\label{neuma}
m_{\n} = -m_D{1\over M}m_D^T \ \sim\ \e^{2a} \left(\begin{array}{ccc}
    \e^2  & \e  & \e \\
    \e  & \; 1 \; & 1 \\
    \e  &  1  & 1 
    \end{array}\right)\;.
\end{equation}
Note, that the structure of this matrix is determined by the 
U(1)$_F$ charges of the $\bf 5^*$-plets only. It is independent of the 
U(1)$_F$ charges of the right-handed neutrinos. 

Since all elements of the 2-3 submatrix of (\ref{neuma}) are ${\cal O}(1)$,
one naturally obtains a large $\n_\m -\n_\t$ mixing angle 
$\Q_{23}$ \cite{sy98,ilr98}. At first sight 
one may expect $\Q_{12} = {\cal O}(\e)$, which would correspond to
the SMA solution of the MSW effect. However, one can also have
a large mixing angle $\Q_{12}$ if the determinant of the 2-3 submatrix
of $m_\n$ is ${\cal O}(\e)$ \cite{vis98}. Choosing the coefficients
${\cal O}(1)$ randomly, in the spirit of `flavour anarchy' \cite{mur02}, 
the SMA and the LMA solutions are about equally probable for 
$\e \simeq 0.1$ \cite{sy00}. 
The corresponding neutrino masses are consistent with
$m_2 \sim 5\times 10^{-3}$~eV and $m_3 \sim 5\times 10^{-2}$~eV. 
We conclude that the neutrino mass matrix (\ref{neuma}) naturally yields a 
large angle $\Q_{23}$, with $\Q_{12}$ being either large or small. 
In order to have maximal mixings the coefficients ${\cal O}(1)$ have to
obey special relations.  

\subsection{Models with SO(10)}

Neutrino masses in SO(10) GUTs have been extensively discussed in the literature
\cite{ref02}. Since all quarks and leptons are unified in a single multiplet
these models often have difficulties to reconcile the large neutrino mixings
with the small quark mixings. In the following we shall illustrate this
problem by means of an example \cite{bw01} which only makes use of the seesaw 
relation, the SO(10) relation between the up-quark and Dirac neutrino mass 
matrices, 
\begin{equation}
m_u = m_D\;,
\end{equation}
and the empirically known properties of the up-quark mass matrix.

With $m_D = m_u$ the seesaw mass relation becomes
\begin{equation}
m_\n \simeq - m_u {1\over M} m_u^T \;.
\end{equation}
The large neutrino mixings now appear very puzzling, since the quark mass 
matrices are hierarchical and the quark mixings are small. It turns out,
however, that because of the known properties of the up-quark mass matrix
this puzzle can be resolved provided the heavy neutrino masses also obey 
a specific hierarchy. This then leads to predictions for a number of 
observables in neutrino physics including the cosmological baryon asymmetry.

From the phenomenology of weak decays we know that the quark matrices have
approximately the form \cite{qm02},
\begin{equation}\label{qmass}
m_{u,d} \propto \left(\begin{array}{ccc}
    0  & \e^3 e^{i\phi}  & 0 \\
    \e^3 e^{i\phi}  & \; \r\e^2 \; & \h\e^2 \\
    0  &  \h\e^2  & e^{i\j} 
    \end{array}\right) \;.
\end{equation}
Here $\e \ll 1$ is the parameter which determines the flavour
mixing, and $\r = |\r| e^{i\a}$, $\h = |\h| e^{i\b}$
are complex parameters ${\cal O}(1)$.
We have chosen a `hierarchical' basis, where off-diagonal
matrix elements are small compared to the product of the corresponding
eigenvalues, $|m_{ij}|^2 \leq {\cal O}(|m_i m_j|)$. In contrast to the 
usual assumption of hermitian mass matrices, 
SO(10) invariance dictates the matrices to be symmetric.
All parameters may take different values
for up$-$ and down$-$quarks. Typical choices for $\e$ are $\e_u \simeq 0.07$, 
$\e_d \simeq 0.2$. The agreement with data can be improved by
adding in the 1-3 element a term ${\cal O}(\e^4)$
which, however, is not important for the following analysis. 
In case of a 1-3 element ${\cal O}(\e^3)$, one can have a SU(3) generation
symmetry leading to different results \cite{kr01}. Data also 
fix one product of phases to be `maximal', i.e. 
$\Delta = \phi_u-\a_u - \phi_d + \a_d \simeq \pi/2$.

We do not know the structure of the Majorana mass matrix $M = h_N \VEV \Phi$.
However, in models with family symmetries it should be similar to the quark mass 
matrices, i.e. the structure should be independent of the Higgs field.
In this case, one expects 
\begin{equation}\label{Mtext}
M = \left(\begin{array}{ccc}
    0  & M_{12}  & 0 \\
    M_{12}  & M_{22} \; & M_{23} \\
    0  &  M_{23}  & M_{33} 
    \end{array}\right) \;,
\end{equation}
with $M_{12} \ll M_{22} \sim M_{23} \ll M_{33}$. The symmetric mass matrix $M$ 
is diagonalized by a unitary matrix, 
\begin{equation}
U^{(N)\dg} M U^{(N)*} = \left(\begin{array}{ccc} 
M_1 & 0 & 0 \\ 0 & M_2 & 0 \\ 0 & 0 & M_3 \end{array}\right).
\end{equation}
Using the seesaw formula
one can now evaluate the light neutrino mass matrix. Since the choice of
the Majorana matrix $M$ fixes a basis for the right-handed neutrinos the
allowed phase redefinitions of the Dirac mass matrix $m_D$ are restricted.
In eq.~(\ref{qmass}) the phases of all matrix elements have therefore been 
kept. 

The $\n_\m$-$\n_\t$ mixing angle is known to be large. This leads us to
require $m_{\n_{i,j}}={\cal O}(1)$ for $i,j =2,3$. It is remarkable that this
determines the hierarchy of the heavy Majorana mass matrix to be
\begin{equation}
M_{12} : M_{22} : M_{33} = \e^5 : \e^4 : 1\;.
\end{equation}  
With $M_{33} \simeq M_3$, $M_{22} = \s \e^4 M_3$, 
$M_{23} = \z \e^4 M_3 \sim M_{22}$ and
$M_{12} = \e^5 M_3$, one obtains for masses and mixings to order 
${\cal O}(\e^4)$,
\begin{equation}\label{Mhier}
M_1 \simeq - {\e^6\over \s}  M_3\;, \quad M_2 \simeq \s \e^4 M_3\;,
\end{equation}
with $U^{(N)}_{12} = - U^{(N)}_{21} = \e/ \s$,
$U^{(N)}_{23} = {\cal O}(\e^4)$ and $U^{(N)}_{13} = 0$.
Note, that $\s$ can always be chosen real.
This yields for the light neutrino mass matrix
\begin{equation}\label{nmass1}
 m_{\n}  = - \left(\begin{array}{ccc}
    0  & \e e^{2i\phi}  & 0\\
    \e e^{2i\phi}   & -\s e^{2i\phi} + 2\r e^{i\phi}   & \h e^{i\phi}  \\
    0  & \h e^{i\phi}   & \; e^{2i\j}
    \end{array}\right)\ {v_1^2 \over M_3}  \;.
\end{equation}
The complex parameter
$\z$ does not enter because of the hierarchy.
The matrix (\ref{nmass1}) has the same structure as the mass matrix 
(\ref{neuma}) in the SU(5)$\times$U(1)$_F$ model, except for additional 
texture zeroes.
Since, as required, all elements of the 2-3 submatrix are ${\cal O}(1)$, 
the mixing angle $\Q_{23}$ is naturally large. A large mixing 
angle $\Q_{12}$ can again occur in case of a small determinant of the
2-3 submatrix.
Such a condition can be fullfilled without fine tuning for $\s, \r, 
\h = {\cal O}(1)$. 

The mass matrix $m_\n$ is again diagonalized by a unitary matrix, 
$U^{(\n)\dg} m_{\n} U^{(\n)*}  = \mbox{diag}(m_1,m_2,m_3)$.
A straightforward calculation yields ($s_{ij} = \sin{\Q_{ij}}$,
$c_{ij} = \cos{\Q_{ij}}$, $\xi=\e/(1+|\h|^2)$),
\begin{equation}\label{nmix}
U^{(\n)}  = \left(\begin{array}{ccc}
 c_{12} e^{i(\phi-\b+\j-\g)}  & s_{12} e^{i(\phi-\b+\j-\g)}  & 
             \xi s_{23} e^{i(\phi-\b+\j)}\\
 - c_{23}s_{12} e^{i(\phi+\b-\j+\g)} & c_{23}c_{12} e^{i(\phi+\b-\j+\g)}  & 
              s_{23} e^{i(\phi+\b-\j)}  \\
  s_{23}s_{12} e^{i(\g+\j)} & -s_{23}c_{12} e^{i(\g+\j)}   & c_{23} e^{i\j}
    \end{array}\right) \;,
\end{equation}
with the mixing angles,
\begin{equation}
\tan{2\Q_{23}} \simeq {2|\h|\over 1-|\h|^2}\;, \quad
\tan{2\Q_{12}} \simeq 2 \sqrt{1+|\h|^2} {\e \over \d}\;.
\end{equation}
Note, that the 1-3 element of the mixing matrix is small, 
$U^{(\n)}_{13} = {\cal O}(\e)$. 
The masses of the light neutrinos are
\beqa
m_1 \simeq  - \tan^2{\Q_{12}}\ m_2\;,\quad
m_2 \simeq {\e \over (1+|\h|^2)^{3/2}} \cot{ \Q_{12}}\ m_3\;,\quad
m_3 \simeq (1+|\h|^2)\ {v_1^2\over M_3}\;.
\eeqa 
This corresponds to the weak hierarchy,
\begin{equation}
m_1 : m_2 : m_3 = \e : \e : 1 \;,
\end{equation} 
with $m_2^2 \sim m_1^2 \sim \D m_{21}^2 = m_2^2-m_1^2 \sim \e^2$. Since
$\e \sim 0.1$, this pattern is consistent with the LMA solution of the
solar neutrino problem, but not with the LOW solution.

The large $\n_\m$-$\n_\t$ mixing is related to the
very large mass hierarchy (\ref{Mhier}) of the heavy Majorana 
neutrinos. The large $\n_e$-$\n_\m$ mixing follows from the
particular values of parameters ${\cal O}(1)$.
Hence, one expects two large mixing angles, but
single maximal or bi-maximal mixing would require fine tuning. 
On the other hand, one definite prediction is the occurence of exactly 
one small matrix element, $U^{(\n)}_{13} = {\cal O}(\e)$. 
Note, that the obtained pattern of neutrino mixings is independent of the 
off-diagonal elements of the mass matrix $M$. For instance, replacing the 
texture (\ref{Mtext}) by a diagonal matrix, $M = \mbox{diag}(M_1,M_2,M_3)$, 
leads to the same pattern of neutrino mixings.

In order to calculate various observables in neutrino physics we need
the leptonic mixing matrix 
\begin{equation}
U = U^{(e)\dg} U^{(\n)}\;,
\end{equation}
where $U^{(e)}$ is the charged lepton mixing matrix. In our framework we
expect $U^{(e)} \simeq V^{(d)}$, and also $V = V^{(u)\dg} V^{(d)}
\simeq V^{(d)}$ for the CKM matrix since $\e_u < \e_d$. This yields
for the leptonic mixing matrix 
\begin{equation}\label{lmix}
U \simeq  V^{\dg} U^{(\n)} \;.
\end{equation}
To leading order in the Cabibbo angle $\l \simeq 0.2$ we only need the
off-diagonal elements $V^{(d)}_{12} = \Bar{\l} = - V^{(d)*}_{21}$. Since
the matrix $m_d$ is complex, the Cabibbo angle is modified by phases,
$\Bar{\l} = \l\exp{\{i(\phi_d-\a_d)\}}$. The resulting leptonic mixing 
matrix is indeed of the wanted form (\ref{umix}) with all matrix elements 
${\cal O}(1)$, except $U_{13}$,
\begin{equation}
U_{13} = \xi s_{23} e^{i(\phi-\b+\j)} - \Bar{\l} s_{23} e^{i(\phi+\b-\j)} 
       =  {\cal O}(\l,\e) \sim 0.1 \;,
\end{equation}
which is close to the experimental limit. 

Let us now consider the CP violation in neutrino oscillations. Observable 
effects are  controlled by the Jarlskog parameter (cf.~(\ref{jarls}))
for which one finds 
\begin{equation}
J_l \simeq {\l\over 4\sqrt{2}} \sin{\d}\;,
\end{equation}
where $\d$ is some function of the unknown parameters $\co(1)$.
Due to the large neutrino mixing angles, $J_l$ can be much bigger than the
Jarlskog parameter in the quark sector, $J_q = {\cal O}(\l^6) \sim 10^{-5}$.

According to the seesaw mechanism neutrinos are Majorana fermions. This
can be directly tested in neutrinoless double $\b$-decay. The decay
amplitude is proportional to the complex mass 
\begin{equation}
\VEV m = \sum_i U_{ei}^2 m_i 
= - {1 \over 1 + |\h|^2} \left(\l^2 |\h|^2 e^{2i(\phi_d -\a_d + \b + \phi- \j)}
                 - 2 \l \e e^{i(\phi_d -\a_d + 2 \phi)}\right) m_3\;. 
\end{equation}
With $m_3 \simeq \sqrt{\D m^2_{atm}} \simeq 5\times 10^{-2}$~eV this yields 
$\VEV m \sim  10^{-3}$~eV, more than two orders of magnitude
below the present experimental upper bound \cite{hei01}.

\section{Leptogenesis}

One of the main successes of the standard early-universe cosmology is the
prediction of the abundances of the light elements, D, $^3$He, $^4$He and 
$^7$Li. Agreement between theory and observation is obtained for
a certain range of the parameter $\eta$, the ratio of baryon density and
photon density \cite{rpp00},
\begin{equation}
\eta = {n_B\over n_\g} = (1.5 - 6.3)\times 10^{-10}\;,
\end{equation}
where the present number density of photons is $n_\g \sim 400/{\rm cm}^3$. 
Since no significant amount of antimatter is observed in the universe, 
the baryon density yields directly the cosmological baryon asymmetry, 
$Y_B =(n_B - n_{\bar{B}})/s \simeq \eta/7$, where $s$ is the entropy density.

A matter-antimatter asymmetry can be dynamically generated in an expanding
universe if the particle interactions and the cosmological evolution satisfy 
Sakharov's conditions \cite{sa67}, 
\begin{itemize}
\item baryon number violation\;,
\item $C$ and $C\!P$ violation\;,
\item deviation from thermal equilibrium .
\end{itemize}
Although the baryon asymmetry is just a single number, it provides an
important relationship between the standard model of cosmology, i.e. the
expanding universe with Robertson-Walker metric, and the standard model
of particle physics as well as its extensions.

At present there exist a number of viable scenarios for baryogenesis. They
can be classified according to the different ways in which Sakharov's 
conditions are realized. In grand unified theories
$B$ and $L$ are broken by the interactions of gauge bosons and leptoquarks.
This is the basis of classical GUT baryogenesis \cite{kt90}.
Analogously, the lepton number violating decays of heavy Majorana neutrinos 
lead to leptogenesis \cite{fy86}. In the simplest version of
leptogenesis the initial abundance of the heavy neutrinos is generated 
by thermal processes. Alternatively,
heavy neutrinos may be produced in inflaton decays, in the reheating process
after inflation, or by brane collisions \cite{all02}. The observed magnitude
of the baryon asymmetry can be obtained for realistic neutrino masses \cite{bp00}.

The crucial deviation from thermal equilibrium can also be realized in several
ways. One possibility is a sufficiently strong first-order electroweak phase 
transition which makes electroweak baryogenesis possible \cite{rt99}. 
For the classical GUT baryogenesis and for leptogenesis the departure from
thermal equilibrium is due to the deviation of the number density of the
decaying heavy particles from the equilibrium number density.
How strong this departure from equilibrium is depends on the lifetime
of the decaying heavy particles and the cosmological evolution. 

\subsection{Baryon and lepton number at high temperatures}

The theory of baryogenesis involves non-perturbative aspects of quantum
field theory and also non-equilibrium statistical field theory, in particular
the theory of phase transitions and kinetic theory. 
\begin{figure}[h]
\begin{center}
\scaleboxto(7,7) {\parbox[c]{9cm}{\input{Sphaleron.tex}}}
\end{center}
\caption{One of the 12-fermion processes which are in thermal 
equilibrium in the high-temperature phase of the standard model.
\label{fig_sphal}}
\end{figure}
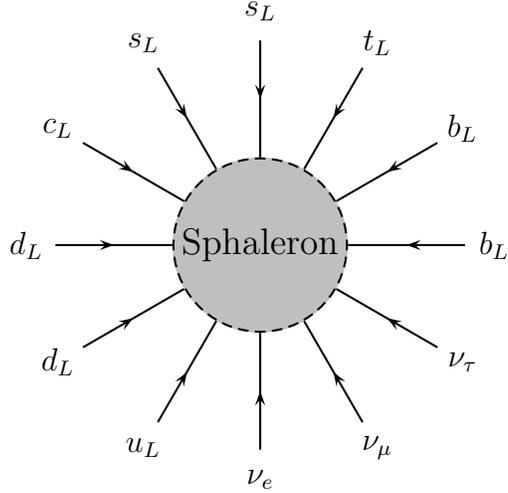
A crucial ingredient is the connection between baryon number and lepton 
number in the high-temperature, symmetric phase of
the standard model. Due to the chiral nature of the weak interactions $B$ and
$L$ are not conserved. At zero temperature this has no observable 
effect due to the smallness of the weak coupling. However, as the temperature 
approaches the critical temperature $T_{EW}$ of the electroweak transition, 
$B$- and $L$-violating processes come into thermal equilibrium \cite{krs85}. 

The rate of these processes is
related to the free energy of sphaleron-type field configurations which carry
topological charge. In the standard model they lead to an effective
interaction of all left-handed quarks and leptons \cite{tho76} 
(cf. fig.~\ref{fig_sphal}), 
\begin{equation}\label{obl}
O_{B+L} = \prod_i \left(q_{Li} q_{Li} q_{Li} l_{Li}\right)\; ,
\end{equation}
which violates baryon and lepton number by three units, 
\begin{equation} 
    \D B = \D L = 3\;. \label{sphal1}
\end{equation}

The evaluation of the sphaleron rate in the symmetric high-temperature phase 
is a complicated problem. A clear physical picture has been obtained in
B\"odeker's effective theory \cite{boe98} according to which low-frequency 
gauge field fluctuations satisfy an equation analogous to electric
and magnetic fields in a superconductor,
\begin{equation}
{\bf D}\times {\bf B} = \s {\bf E} - {\bf \z}\; .
\end{equation}
Here $\bf \z$ represents Gaussian noise, and $\s$ is a non-abelian 
conductivity. The sphaleron rate can then be written as,
\begin{equation}
\G_{SPH} \simeq (14.0 \pm 0.3) {1\over \s} (\a_w T)^5\;.
\end{equation}
Lattice simulations \cite{bmr00} have confirmed early estimates that $B$- and 
$L$-violating processes are in thermal equilibrium for temperatures in the 
range
\begin{equation}
T_{EW} \sim 100\ \mbox{GeV} < T < T_{SPH} \sim 10^{12}\ \mbox{GeV}\;.
\end{equation}

Sphaleron processes have a profound effect on the generation of the
cosmological baryon asymmetry, in particular in connection with the dominant
lepton number violating interactions between lepton and Higgs fields,
\begin{equation}\label{dl2}
\cl_{\Delta L=2} ={1\over 2} f_{ij}\ l^T_{Li}\f C l_{Lj}\f 
                  +\mbox{ h.c.}\;.
\end{equation}
As discussed in section~3.3, these interactions arise from the exchange of 
heavy Majorana neutrinos, with
$\f = H_1$ and $f_{ij} = -(h_\n M^{-1} h^T_\n)_{ij}$.
In the Higgs phase of the standard model, where the Higgs field acquires a 
vacuum expectation value, the interaction (\ref{dl2}) leads to
Majorana masses for the light neutrinos $\n_e$, $\n_\m$ and $\n_\t$.   

One may be tempted to conclude from eq.~(\ref{sphal1}) that any
$B+L$ asymmetry generated before the electroweak phase transition,
i.e. at temperatures $T>T_{EW}$, will be washed out. However, since
only left-handed fields couple to sphalerons, a non-zero value of
$B+L$ can persist in the high-temperature, symmetric phase in case
of a non-vanishing $B-L$ asymmetry. An analysis of the chemical potentials
of all particle species in the high-temperature phase yields a
relation between the baryon asymmetry $Y_B = (n_B-n_{\bar{B}})/s$ and the 
corresponding $B-L$ and $L$ asymmetries 
$Y_{B-L}$ and $Y_L$, respectively,
\begin{equation}\label{basic}
Y_B\ =\ a\ Y_{B-L}\ =\ {a\over a-1}\ Y_L\;.
\end{equation}
The number $a$ depends on the other processes which are in thermal 
equilibrium. If these are all standard model interactions one has 
$a=8/23$ in the case of two Higgs doublets \cite{ks88}. 

From eq.~(\ref{basic}) one concludes that the cosmological baryon asymmetry
requires also a lepton asymmetry, and therefore lepton number violation.
This leads to an intriguing interplay
between Majorana neutrinos masses, which are generated by the lepton-Higgs
interactions (\ref{dl2}), and the baryon asymmetry:
lepton number violating interactions are needed in order to generate a
baryon asymmetry; however, they have to be sufficiently weak, so that they
fall out of thermal equilibrium at the right time and a generated asymmetry
can survive until today.

\subsection{Thermal leptogenesis}

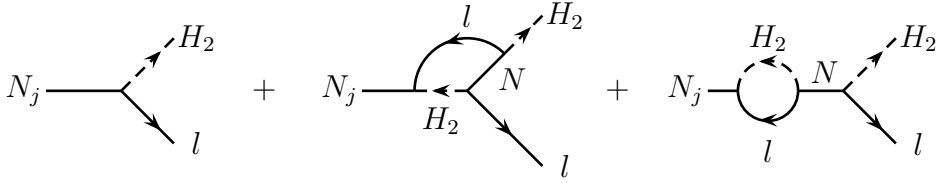
\begin{figure}
\centerline{\input{Decay.tex}}
\caption{\it Tree level and one-loop diagrams contributing to heavy
    neutrino decays. \label{fig:decay}}
\end{figure}

Let us now consider the simplest possibility for a departure from thermal
equilibrium, the decay of heavy, weakly interacting particles in a thermal
bath. We choose the heavy particle to be the lightest of the heavy
Majorana neutrinos, $N_1 \equiv N =N^c$, which can decay into a lepton Higgs  
pair $l \phi$ and also into the $C\!P$ conjugate state $\bar{l} \bar{\phi}$,
\begin{equation}
N \rightarrow l\;\phi\;, \quad N \rightarrow \bar{l}\;\bar{\phi}\;.
\end{equation}
In the case of $C\!P$ violating couplings a lepton asymmetry can be generated in
the decays of the heavy neutrinos $N$,
\begin{equation}\label{gcp} 
\G(N\rightarrow l\phi) = {1\over 2}(1+\ve_1)\G\;,\quad
\G(N\rightarrow \bar{l}\bar{\phi})={1\over 2}(1-\ve_1)\G\;,
\end{equation}
where $\G$ is the total decay width and $\ve_1 \ll 1$ measures the amount of
$C\!P$ violation.

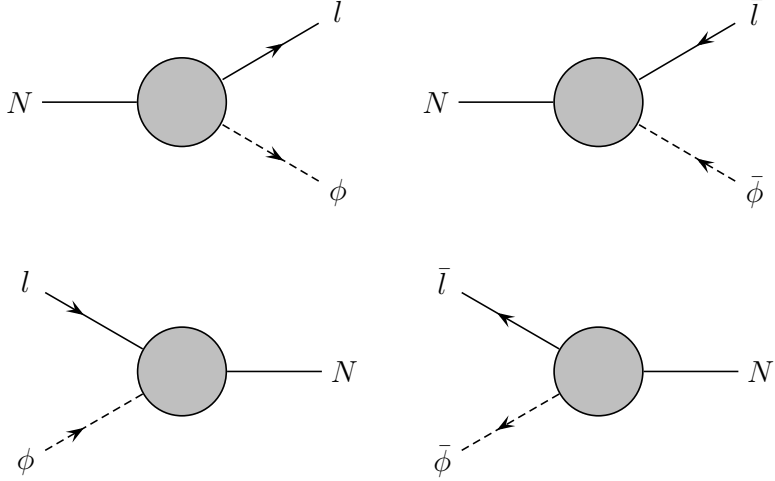
\begin{figure}[h]
\begin{center}
\scaleboxto(5.4,0){\parbox[c]{9cm}{\input{decay1.tex}}}
\scaleboxto(5.4,0){\parbox[c]{9cm}{\input{decay2.tex}}}
\scaleboxto(5.4,0){\parbox[c]{9cm}{\input{decay3.tex}}}
\scaleboxto(5.4,0){\parbox[c]{9cm}{\input{decay4.tex}}}
\end{center}
\caption{\it $\D L=1$ processes: decays and inverse decays of a heavy Majorana
neutrino.\label{decinv}}
\end{figure}

The $C\!P$ asymmetry $\ve_1$ arises from one-loop vertex and self-energy 
corrections (fig.~\ref{fig:decay}) \cite{fps95,crv96,bp98}. It can be expressed 
in a compact form, which in the mass eigenstate basis of the heavy neutrinos
reads,
\begin{equation}\label{cpa}
\ve_1 \simeq {3\over 16\pi} \mbox{sign}(M_1) 
{\mbox{Im}\left(m_D^\dg m_\n m_D^*\right)_{11} 
 \over v_1^2 \wt{m}_1}\;.
\end{equation}
Here, in addition to the mass matrices $m_D$ and $m_\n$, an effective neutrino mass,
\begin{equation}
\wt{m}_1 = {\left(m_D^\dg m_D\right)_{11} \over |M_1|}\;,
\end{equation}
appears, which is a sensitive parameter for successful leptogenesis \cite{plu97}.
Note, that the maximal $C\!P$ asymmetry is related to the mass $M_1$ of the 
heavy Majorana neutrino $N_1$ \cite{di02}. In the case of mass differences
of order the decay widths, $|M_i - M_1| = {\cal O}(\G_i,\G_1)$, $i\neq 1$,
the $C\!P$ asymmetry is enhanced \cite{pil99}. Contrary to the case considered
here, decays of $N_2$ or $N_3$ may be the origin of the baryon asymmetry
\cite{nt01}, if washout effects are sufficiently small.

\begin{figure}[h]
\begin{center}
\scaleboxto(5.4,0){
\parbox[c]{9cm}{\input{twotwo.tex}}}
\scaleboxto(5.4,0){
\parbox[c]{9cm}{\input{twoanti.tex}}}
\end{center}
\caption{\it $\D L=0$ and $\D L=2$ lepton Higgs processes.\label{lephig}}
\end{figure}
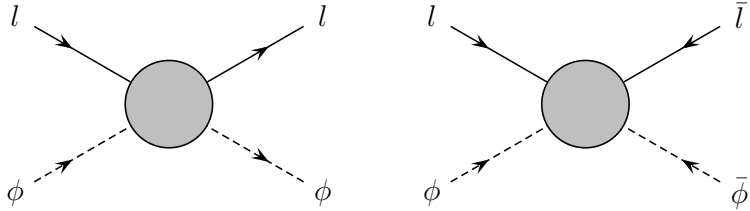

The generation of a baryon asymmetry is an out-of-equilibrium process 
which is generally treated by means of Boltzmann equations \cite{kt90}. 
The main processes in the thermal bath are the decays
and the inverse decays of the heavy neutrinos (fig.~\ref{decinv}), and
the lepton number conserving ($\D L=0$) and violating ($\D L=2$) processes
(fig.~\ref{lephig}). In addition there are other processes, in particular
those involving the t-quark, which are important in a quantitative
analysis \cite{lut92,plu97}. 

\begin{figure}
\mbox{ }\hfill
\epsfig{file=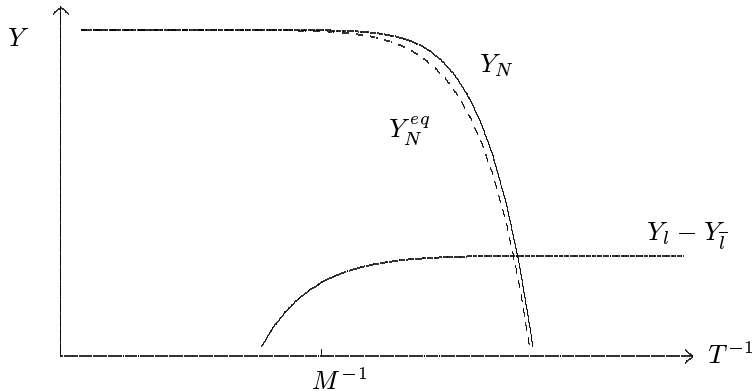,width=10cm}
\hfill\mbox{ }
\caption{\label{fig:outofequ}\it Time evolution of the number density to 
entropy density ratio. At $T\sim M$ the system gets out of
equilibrium and an asymmetry is produced.}
\end{figure}

A typical solution of the Boltzmann equations is
shown in fig.~\ref{fig:outofequ}. Here the ratios of number densities and
entropy density,
\begin{equation}
Y_X = {n_X\over s}\;,
\end{equation}
are plotted, which remain constant for an expanding universe in thermal 
equilibrium. A heavy neutrino, which is weakly coupled 
to the thermal bath, falls out of thermal equilibrium at temperatures 
$T \sim M$, since its decay is too slow to follow the rapidly decreasing 
equilibrium distribution $f_N \sim \exp(-\b M)$. This leads to an excess of 
the number density, $n_N > n_N^{eq}$. $C\!P$ violating partial decay widths 
then yield a lepton
asymmetry which, by means of sphaleron processes, is partially transformed
into a baryon asymmetry.

The $C\!P$ asymmetry $\ve_1$ (\ref{cpa}) leads to a lepton asymmetry in the
course of the cosmological evolution, which is then partially transformed
into a baryon asymmetry \cite{fy86} by sphaleron processes,
\begin{equation}\label{basym}
Y_B\ =\ {n_L-n_{\Bar{L}}\over s}\ =\ c_S\k\ {\ve_1\over g_*}\;,
\end{equation}
where $c_S = -8/15$ is the sphaleron conversion factor in the case of two Higgs 
doublets. In order to determine the washout factor $\k<1$ one has to solve the 
Boltzmann equations.
Compared to other scenarios of baryogenesis this leptogenesis mechanism
has the advantage that, at least in principle, the resulting baryon asymmetry
is entirely determined by neutrino properties. 

We are now ready to determine the baryon asymmetries predicted by the models
of neutrino masses discussed in section~5. 
Since for the Yukawa couplings only the powers in $\e$ are known, we will
also obtain the $C\!P$ asymmetry and the corresponding baryon asymmetry
to leading order in $\e$, i.e. up to unknown factors ${\cal O}(1)$. 

Consider first the  model with symmetry group SU(5)$\times$U(1)$_F$ 
(cf.~section~5.2). One easily obtains from eqs.~(\ref{yuk}), (\ref{cpa}) 
and table~\ref{tab:lop},
\begin{equation}
  \ve_1\ \sim\ {3\over 16\pi}\ \e^4\;.
  \label{su5epsilon}
\end{equation} 
With $\e^2 \sim 1/300$  and $g_* \sim 100$ this yields the baryon asymmetry,
\begin{equation}\label{est1}
Y_B \sim \k\ 10^{-8}\;.
\end{equation}
For $\k \sim 0.1\ldots 0.01$ this is indeed the correct order of magnitude.
The baryogenesis temperature is given by the mass of the lightest of the
heavy Majorana neutrinos,
\begin{equation}
T_B \sim M_1 \sim \e^4 M_3 \sim 10^{10}\ \mbox{GeV}\;.
\end{equation}
This is essentially the model studied in ref.~\cite{bp96}, where the 
$C\!P$ asymmetry is determined by the mass hierarchy of light and heavy Majorana 
neutrinos. It is remarkable that the observed
baryon asymmety is obtained without any fine tuning of parameters, if
$B-L$ is broken at the unification scale $\L_{GUT}$. Note, that
the generated baryon asymmetry does not depend on the flavour mixing of the 
light neutrinos.

\begin{figure}
\mbox{ }\hfill
\epsfig{file=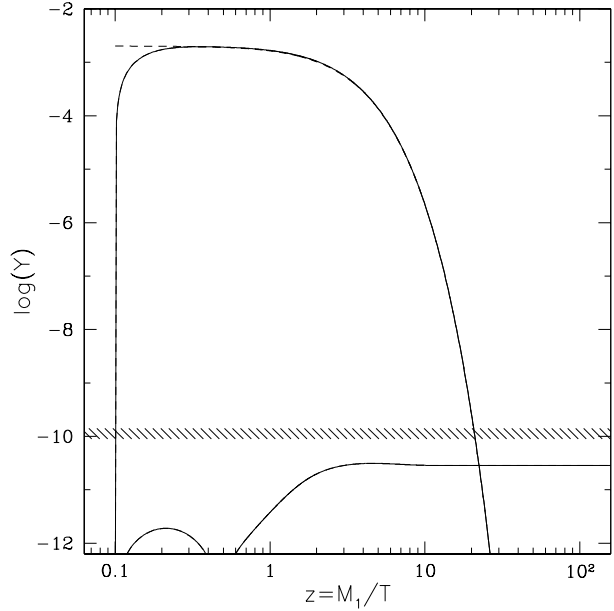,width=8.2cm}
\hfill\mbox{ }
\caption{\it Time evolution of the heavy neutrino number density and the
lepton asymmetry for the SO(10) model. The upper solid line shows the solution of 
the Boltzmann equations for the right-handed neutrinos; the  
equilibrium distribution is represented by the dashed line. The absolute value of 
the lepton asymmetry $Y_L$ is given by the lower solid line, and the hatched area 
shows the lepton asymmetry corresponding to the observed baryon asymmetry 
\cite{bjp02}.\label{fig:asy1}}
\end{figure} 
\begin{figure}
\mbox{ }\hfill
\epsfig{file=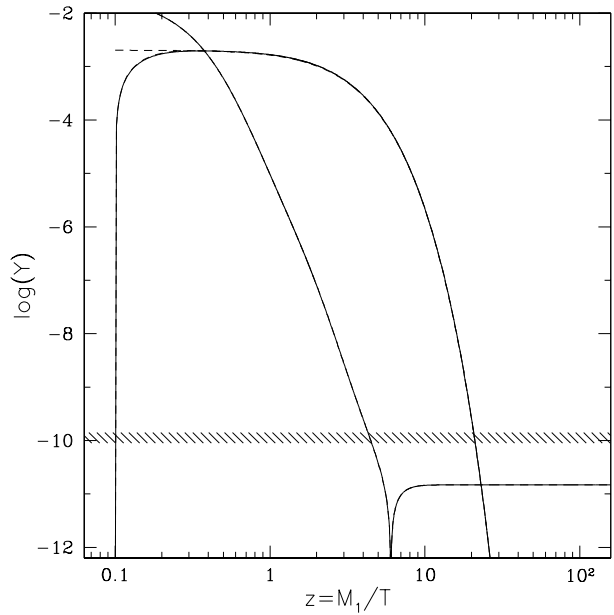,width=8.2cm}
\hfill\mbox{ }
\caption{Same as fig.~\ref{fig:asy1} with initial asymmetry $Y_L^{in} \simeq 10^{-2}$
\cite{bjp02}.\label{fig:asy2}}
\end{figure} 

Qualitatively, the SO(10) model discussed in section 5.3 is very similar. For a 
class of parameters corresponding to the LMA MSW-solution with $\s,\r,\h = \co(1)$,
one finds for the $C\!P$ asymmetry \cite{bw01},
\begin{equation}
\ve_1 \simeq {3\over {16\pi}} \e^6 {|\h|^2\over \s}  
         {(1+|\r|)^2 \over |\h|^2 +|\r|^2}\sin(\phi_u-\a_u) \;.
\end{equation}  
Note, that except for $\s$, the $C\!P$ asymmetry is determined by parameters
of the quark mass matrices. Similar results have been obtained in 
refs.~\cite{jpr01,bmx01} for different models. For a recent analysis within SO(10)
models, which favours the Just-So and SMA solar neutrino solutions, see ref.~\cite{bg02}.

Numerically, with $\e \sim 0.1$ one finds $\ve_1 \sim 10^{-7}$ and
\begin{equation}
|M_1| \simeq e^6 {(1+|\h|^2) v_1^2\over \s m_3} \sim 10^9\ \mbox{GeV}\;,
\end{equation}  
with
$\wt{m}_1 \sim (|\h|^2+|\r|^2) m_3/(\s(1+|\h|^2)) \sim 10^{-2}$~eV. 
The baryon asymmetry is then given by
\begin{equation}\label{asym}
Y_B \sim  - \k~\mbox{sign}(\s)~\sin{(\phi_u-\a_u)} \times 10^{-9}\;.
\end{equation}
The solution of the full Boltzmann equations is shown in fig.~\ref{fig:asy1}. 
The initial condition at a temperature $T \sim 10 M_1$ is 
chosen to be a state without heavy neutrinos and with vanishing lepton asymmetry.
The Yukawa interactions are sufficient to bring the heavy neutrinos into thermal 
equilibrium. At temperatures $T\sim M_1$ the familiar out-of-equilibrium decays set in,
which leads to a non-vanishing baryon asymmetry. The dip in 
fig.~\ref{fig:asy1} is due to a change of sign in the lepton asymmetry at 
$T \sim M_1$. The final asymmetry is about 1/3 of the observed value, which lies 
within the present range of theoretical uncertainties. 

A very important question for leptogenesis, and baryogenesis in general, is the
dependence on initial conditions. As fig.~\ref{fig:asy1} demonstrates, the
heavy neutrinos are initially indeed in thermal equilibrium. One may also wonder,
how sensitive the final lepton asymmetry is to an initial asymmetry which may 
have been
generated by some other mechanism. The SO(10) model under consideration turns out
to be very efficient in establishing a symmetric initial state. This can be seen
in fig.~\ref{fig:asy2} where the maximal asymmetry $Y_L^{in} \simeq 10^{-2}$ has
been assumed as initial condition. Within one order of magnitude in
temperature this initial asymmetry is washed out by eight orders of magnitude!
Hence, the final baryon asymmetry is a definite prediction of the theory, 
independent of initial conditions.

In summary, the experimental evidence for small neutrino masses and
large neutrino mixings, together with the known small quark mixings, have
important implications for the structure of grand unified theories.
In SU(5) models this
difference between the lepton and quark sectors can be explained by 
U(1)$_F$ family symmetries. In these models the heavy Majorana neutrino masses
are not constrained by low energy physics, i.e. light neutrino masses and
mixings. Successful leptogenesis is possible, but it depends on the choice of the 
heavy Majorana neutrino masses.

In SO(10) models the implications of large neutrino mixings are much more
stringent because of the connection between Dirac neutrino and up-quark mass
matrices. The requirement of large neutrino mixings then
determines the relative magnitude of the heavy Majorana neutrino masses in 
terms of the known quark mass hierarchy. This leads to predictions
for neutrino mixings and masses, $C\!P$ violation in neutrino oscillations and
neutrinoless double $\beta$-decay. It is remarkable that the predicted order of 
magnitude of the baryon asymmetry is also in accord with observation.\\ 

\noindent
I would like to thank the participants of the school for stimulating questions
and the organizers for arranging an enjoyable and fruitful meeting at
Beatenberg. I have also benefited from discussions with P.~Di~Bari, M.~Pl\"umacher,
A.~Ringwald, S.~Stodolsky and D.~Wyler, and I thank S.~Wiesenfeldt for providing
some of the figures.

\end{document}

%% file: kurie.tex
\begin{picture}(360,220)(0,0)
  \SetOffset(40,30)
  \LongArrow(0,0)(330,0) \Text(345,0)[]{$E_e$}
  \LongArrow(0,0)(0,170) \Text(-30,160)[]{$K(E_e)$}
  \SetScale{100.} \SetWidth{0.005}
  \DashCurve{(1.8,.6)(1.9,.55)(2,.485)(2.1,.405)(2.15,.34)(2.2,.26)
    (2.22, .192938)(2.24, 0.)}{0.02}
  \Curve{(.0,1.5)(1.8,.6)(3,0)}
  \Text(255,40)[]{$m_\nu=0$} \Text(195,12)[]{$m_\nu\not=0$\,?}
  \Line(3,.0)(3,-.05) \Text(300,-15)[]{$E_0$}
  \Line(2.2385,.01)(2.2385,-.05) \Text(224,-15)[]{$E_0-m_\nu$}
\end{picture}

%% file: doubb.tex
\begin{picture}(170,150)(0,0)
  \SetOffset(0,20)
  \ArrowLine(30,30)(70,30) \Text(10,30)[]{$n$}
  \ArrowLine(70,30)(110,10) \Text(120,5)[]{$p$}
  \Photon(70,30)(110,40){3}{5.5} \Text(88,45)[]{$W$}
  \Vertex(70,30)1 \Vertex(110,40)1
  \ArrowLine(30,80)(70,80) \Text(10,80)[]{$n$}
  \ArrowLine(70,80)(110,100) \Text(120,105)[]{$p$}
  \Photon(110,70)(70,80){3}{5.5} \Text(88,65)[]{$W$}
  \Vertex(70,80)1 \Vertex(110,70)1
  \ArrowLine(110,55)(110,40) \ArrowLine(110,55)(110,70)
  \Line(107,52)(113,58) \Line(113,52)(107,58) \Text(120,55)[]{$\nu$}
  \ArrowLine(110,40)(150,40) \Text(165,40)[]{$e^-$}
  \ArrowLine(110,70)(150,70) \Text(165,70)[]{$e^-$}
\end{picture}
\begin{picture}(70,120)(0,0)
  \SetOffset(0,20)
  \Text(30,55)[]{$\propto m_{ee}$}
\end{picture}

%% file: burst.tex
\begin{picture}(200,100)(0,0)
  \SetOffset(0,20)
  \ArrowLine(30,0)(60,30) \ArrowLine(30,60)(60,30)
  \Text(10,0)[]{$\nu_{\text{\sc relic}}$}
  \Text(10,60)[]{$\nu_{\text{\sc uhe}}$}
  \Photon(60,30)(110,30){4}{6} \Text(90,43)[]{$Z$}
  \Vertex(60,30)1 \Vertex(110,30)1
  \Line(110,30)(140,45) \Line(110,30)(140,15)
  \ArrowLine(144,40)(170,40) \ArrowLine(144,30)(170,30)
  \ArrowLine(144,20)(170,20)
  \GOval(140,30)(15,7)(0)1
  \Text(180,30)[l]{hadrons}
\end{picture}

%% file: nuosc.tex
\begin{picture}(300,100)(0,20)
  \Line(30,51)(60,51) \Line(30,49)(60,49) \Text(15,50)[]{$p$}
  \Line(60,51)(90,71) \Line(60,49)(90,69) \Text(100,75)[]{$n$}
  \ArrowLine(90,30)(60,50) \Text(100,25)[]{$e^+_R$}
  \ArrowLine(65,50)(90,50) \Text(100,50)[]{$\nu_i$}
  \GCirc(60,50){5}{.7} \Text(60,30)[]{$U_{ei}^*$}
  \DashLine(110,50)(185,50)5
  \ArrowLine(210,50)(240,50) \Text(200,50)[]{$\nu_i$}
  \ArrowLine(240,50)(270,30) \Text(280,25)[]{$e^-_L$}
  \ArrowLine(270,70)(240,50) \Text(280,75)[]{$e^+_R$}
  \ArrowLine(245,50)(270,50) \Text(280,50)[]{$\nu_e$}
  \GCirc(240,50){5}{.7}
\end{picture}

%% file: numat.tex
\begin{picture}(100,110)(0,0)
  \ArrowLine(0,0)(50,10) \ArrowLine(50,10)(100,0)
  \Text(0,10)[]{$e$} \Text(100,10)[]{$\nu_e$}
  \ArrowLine(0,60)(50,50) \ArrowLine(50,50)(100,60)
  \Text(0,50)[]{$\nu_e$} \Text(100,50)[]{$e$}
  \Photon(50,10)(50,50){4}{6} \Text(65,30)[]{$W$}
  \Vertex(50,10)1 \Vertex(50,50)1
\end{picture}
\begin{picture}(60,80)(0,20)
\end{picture}
\begin{picture}(100,100)(0,0)
  \ArrowLine(0,0)(50,10) \ArrowLine(50,10)(100,0)
  \Text(0,10)[]{$p,n,e$} \Text(100,10)[]{$p,n,e$}
  \ArrowLine(0,60)(50,50) \ArrowLine(50,50)(100,60)
  \Text(0,45)[]{$\nu_e,\nu_\mu,\nu_\tau$}
  \Text(100,45)[]{$\nu_e,\nu_\mu,\nu_\tau$}
  \Photon(50,10)(50,50){4}{6} \Text(65,30)[]{$Z$}
  \Vertex(50,10)1 \Vertex(50,50)1
\end{picture}

%% file: msw.tex
\begin{picture}(300,230)(0,0)
  \SetOffset(70,50)
  \LongArrow(0,0)(220,0) \Text(235,0)[]{$n_e$}
  \LongArrow(0,0)(0,160) \Text(-65,140)[l]{momentum}
  \ArrowLine(200,10)(160,36) \ArrowLine(160,36)(128,56.5)
  \LongArrow(128,56.5)(108,69) 
  \Curve{(108,69)(96,75.7)(85,78)}
  \ArrowLine(85,78)(70,78) \LongArrow(70,78)(50,78)
  \ArrowLine(50,78)(0,78)
  \Text(180,10)[]{$\nu_e$} \Text(20,70)[]{$\nu_\mu$}
  \DashLine(0,85)(200,85){5}
  \DashLine(0,150)(200,20){6}
  \Line(200,0)(200,160)
  \Text(0,-15)[]{surface} \Text(0,-25)[]{of sun}
  \Text(200,-15)[]{center} \Text(200,-25)[]{of sun}
  \LongArrow(100,-15)(100,-5) \Text(100,-25)[]{$n_e(z_R)$}
\end{picture}

%% file: el.tex
\begin{picture}(100,100)(0,0)
  \SetOffset(0,20)
  \ArrowLine(0,0)(50,10) \ArrowLine(50,10)(100,0)
  \Text(0,10)[]{$e$} \Text(100,10)[]{$\nu_e$}
  \ArrowLine(0,60)(50,50) \ArrowLine(50,50)(100,60)
  \Text(0,50)[]{$\nu_e$} \Text(100,50)[]{$e$}
  \Photon(50,10)(50,50){4}{6} \Text(65,30)[]{$W$}
  \Vertex(50,10)1 \Vertex(50,50)1
\end{picture}
\begin{picture}(70,80)(0,0)
  \SetOffset(0,20)
  \Text(30,30)[]{\Large $+$}
\end{picture}
\begin{picture}(100,100)(0,0)
  \SetOffset(0,20)
  \ArrowLine(0,0)(50,10) \ArrowLine(50,10)(100,0)
  \Text(0,10)[]{$e$} \Text(100,10)[]{$e$}
  \ArrowLine(0,60)(50,50) \ArrowLine(50,50)(100,60)
  \Text(0,45)[]{$\nu_e,\nu_\mu,\nu_\tau$}
  \Text(100,45)[]{$\nu_e,\nu_\mu,\nu_\tau$}
  \Photon(50,10)(50,50){4}{6} \Text(65,30)[]{$Z$}
  \Vertex(50,10)1 \Vertex(50,50)1
\end{picture}

%% file: ano.tex
\begin{picture}(200,100)(0,0)
  \SetOffset(0,20)
  \Line(27,33)(33,27) \Line(27,27)(33,33)
  \ArrowLine(30,30)(120,60) \ArrowLine(120,60)(120,0)
  \ArrowLine(120,0)(30,30)
  \Text(10,30)[]{$J_\mu^{B-L}$}
  \Photon(120,60)(150,60){4}{4} \Text(165,60)[]{$W$}
  \Photon(120,0)(150,0){4}{4} \Text(165,0)[]{$W$}
  \Vertex(120,60)1 \Vertex(120,0)1
\end{picture}

%% file: uni.tex
\begin{picture}(300,200)(0,0)
  \SetOffset(40,30)
  \Curve{(20,116)(200,48)} \Text(100,96)[]{$\alpha_1$}
  \Text(245,100)[]{$U(1)$}
  \Curve{(20,58)(200,48)} \Text(100,60)[]{$\alpha_2$}
  \Text(245,60)[]{$SU(2)$}
  \Curve{(20,16)(200,48)} \Text(100,20)[]{$\alpha_3$}
  \Text(245,25)[]{$SU(3)$}
  \LongArrow(0,0)(230,0) \Text(245,0)[]{$\mu$}
  \LongArrow(0,0)(0,140)
  \Text(-30,130)[]{$\dfrac{1}{\alpha_i(\mu)}$}
  \Line(20,5)(20,-5)
  \Text(10,-15)[l]{$\Lambda_{\text{SUSY}}\sim1$\,TeV}
  \Line(200,5)(200,-5) \Text(200,-15)[]{$\Lambda_{\text{GUT}}$}
\end{picture}

%% file: grav.tex
\begin{picture}(220,100)(0,0)
  \SetOffset(40,20)
  \ArrowLine(30,30)(120,60) \ArrowLine(120,60)(120,0)
  \ArrowLine(120,0)(30,30)
  \Text(60,5)[]{quarks,} \Text(75,-10)[]{leptons}
  \Photon(0,30)(30,30){4}{4} \Text(-25,30)[]{$A_\mu^{(B-L)}$}
  \Photon(120,60)(150,60){4}{4} \Text(170,60)[]{$h_{\mu\nu}$}
  \Photon(120,0)(150,0){4}{4} \Text(170,0)[]{$h_{\mu\nu}$}
  \Vertex(120,60)1 \Vertex(120,0)1 \Text(170,30)[l]{gravitons}
\end{picture}

%% file: Sphaleron.tex
 \begin{center}
     \pspicture*(-0.50,-2.5)(8.5,6.5)
     \psset{linecolor=lightgray}
     \qdisk(4,2){1.5cm}
     \psset{linecolor=black}
     \pscircle[linewidth=1pt,linestyle=dashed](4,2){1.5cm}
     \rput[cc]{0}(4,2){\scalebox{1.5}{Sphaleron}}
     \psline[linewidth=1pt](5.50,2.00)(7.50,2.00)
     \psline[linewidth=1pt](5.30,2.75)(7.03,3.75)
     \psline[linewidth=1pt](4.75,3.30)(5.75,5.03)
     \psline[linewidth=1pt](4.00,3.50)(4.00,5.50)
     \psline[linewidth=1pt](3.25,3.30)(2.25,5.03)
     \psline[linewidth=1pt](2.70,2.75)(0.97,3.75)
     \psline[linewidth=1pt](2.50,2.00)(0.50,2.00)
     \psline[linewidth=1pt](2.70,1.25)(0.97,0.25)
     \psline[linewidth=1pt](3.25,0.70)(2.25,-1.03)
     \psline[linewidth=1pt](4.00,0.50)(4.00,-1.50)
     \psline[linewidth=1pt](4.75,0.70)(5.75,-1.03)
     \psline[linewidth=1pt](5.30,1.25)(7.03,0.25)
     \psline[linewidth=1pt]{<-}(6.50,2.00)(6.60,2.00)
     \psline[linewidth=1pt]{<-}(6.17,3.25)(6.25,3.30)
     \psline[linewidth=1pt]{<-}(5.25,4.17)(5.30,4.25)
     \psline[linewidth=1pt]{<-}(4.00,4.50)(4.00,4.60)
     \psline[linewidth=1pt]{<-}(2.75,4.17)(2.70,4.25)
     \psline[linewidth=1pt]{<-}(1.83,3.25)(1.75,3.30)
     \psline[linewidth=1pt]{<-}(1.50,2.00)(1.40,2.00)
     \psline[linewidth=1pt]{<-}(1.83,0.75)(1.75,0.70)
     \psline[linewidth=1pt]{<-}(2.75,-0.17)(2.70,-0.25)
     \psline[linewidth=1pt]{<-}(4.00,-0.50)(4.00,-0.60)
     \psline[linewidth=1pt]{<-}(5.25,-0.17)(5.30,-0.25)
     \psline[linewidth=1pt]{<-}(6.17,0.75)(6.25,0.70)
     \rput[cc]{0}(8.00,2.00){\scalebox{1.3}{$b_L$}}
     \rput[cc]{0}(7.46,4.00){\scalebox{1.3}{$b_L$}}
     \rput[cc]{0}(6.00,5.46){\scalebox{1.3}{$t_L$}}
     \rput[cc]{0}(4.00,6.00){\scalebox{1.3}{$s_L$}}
     \rput[cc]{0}(2.00,5.46){\scalebox{1.3}{$s_L$}}
     \rput[cc]{0}(0.54,4.00){\scalebox{1.3}{$c_L$}}
     \rput[cc]{0}(0.00,2.00){\scalebox{1.3}{$d_L$}}
     \rput[cc]{0}(0.54,0.00){\scalebox{1.3}{$d_L$}}
     \rput[cc]{0}(2.00,-1.46){\scalebox{1.3}{$u_L$}}
     \rput[cc]{0}(4.00,-2.00){\scalebox{1.3}{$\nu_e$}}
     \rput[cc]{0}(6.00,-1.46){\scalebox{1.3}{$\nu_{\mu}$}}
     \rput[cc]{0}(7.46,0.00){\scalebox{1.3}{$\nu_{\tau}$}}
     \endpspicture
\end{center}

%% file: Decay.tex
\parbox[c]{12.5cm}{
\pspicture(0,0)(3.7,2.6)
\psline[linewidth=1pt](0.6,1.3)(1.6,1.3)
\psline[linewidth=1pt](1.6,1.3)(2.3,0.6)
\psline[linewidth=1pt,linestyle=dashed](1.6,1.3)(2.3,2.0)
\psline[linewidth=1pt]{->}(2.03,0.87)(2.13,0.77)
\psline[linewidth=1pt]{->}(2.03,1.73)(2.13,1.83)
\rput[cc]{0}(0.3,1.3){$N_j$}
\rput[cc]{0}(2.6,0.6){$l$}
\rput[cc]{0}(2.6,2.0){$H_2$}
\rput[cc]{0}(3.5,1.3){$+$}
\endpspicture
\pspicture(-0.5,0)(4.2,2.6)
\psline[linewidth=1pt](0.6,1.3)(1.3,1.3)
\psline[linewidth=1pt,linestyle=dashed](1.3,1.3)(2.0,1.3)
\psline[linewidth=1pt](2,1.3)(2.5,1.8)
\psline[linewidth=1pt,linestyle=dashed](2.5,1.8)(3,2.3)
\psline[linewidth=1pt](2,1.3)(3,0.3)
\psarc[linewidth=1pt](2,1.3){0.7}{45}{180}
\psline[linewidth=1pt]{<-}(1.53,1.3)(1.63,1.3)
\psline[linewidth=1pt]{<-}(1.7,1.93)(1.8,1.96)
\psline[linewidth=1pt]{->}(2.75,2.05)(2.85,2.15)
\psline[linewidth=1pt]{->}(2.5,0.8)(2.6,0.7)
\rput[cc]{0}(0.3,1.3){$N_j$}
\rput[cc]{0}(1.65,0.9){$H_2$}
\rput[cc]{0}(2,2.3){$l$}
\rput[cc]{0}(2.6,1.45){$N$}
\rput[cc]{0}(3.3,2.3){$H_2$}
\rput[cc]{0}(3.3,0.3){$l$}
\rput[cc]{0}(4.0,1.3){$+$}
\endpspicture
\pspicture(-0.5,0)(3.5,2.6)
\psline[linewidth=1pt](0.5,1.3)(0.9,1.3)
\psline[linewidth=1pt](1.7,1.3)(2.3,1.3)
\psarc[linewidth=1pt](1.3,1.3){0.4}{-180}{0}
\psarc[linewidth=1pt,linestyle=dashed](1.3,1.3){0.4}{0}{180}
\psline[linewidth=1pt]{<-}(1.18,1.69)(1.28,1.69)
\psline[linewidth=1pt]{<-}(1.18,0.91)(1.28,0.91)
\psline[linewidth=1pt](2.3,1.3)(3.0,0.6)
\psline[linewidth=1pt,linestyle=dashed](2.3,1.3)(3.0,2.0)
\psline[linewidth=1pt]{->}(2.73,0.87)(2.83,0.77)
\psline[linewidth=1pt]{->}(2.73,1.73)(2.83,1.83)
\rput[cc]{0}(1.3,0.5){$l$}
\rput[cc]{0}(1.3,2){$H_2$}
\rput[cc]{0}(2.05,1.55){$N$}
\rput[cc]{0}(0.2,1.3){$N_j$}
\rput[cc]{0}(3.3,0.6){$l$}
\rput[cc]{0}(3.3,2.0){$H_2$}
\endpspicture
}

%% file: decay1.tex
 \begin{center}
     \pspicture*(-0.5,-0.5)(8.5,4.5)
     \psset{linecolor=lightgray}
     \qdisk(4,2){1cm}
     \psset{linecolor=black}
     \pscircle[linewidth=1pt](4,2){1cm}
     \psline[linewidth=1pt](4.9,2.5)(7.03,3.75)
     \psline[linewidth=1pt](3.00,2.00)(0.90,2.00)
     \psline[linewidth=1pt,linestyle=dashed](4.9,1.5)(7.03,0.25)
     \psline[linewidth=2pt]{->}(6.17,3.25)(6.25,3.30)
     \psline[linewidth=2pt]{->}(6.17,0.75)(6.25,0.70)
     \rput[cc]{0}(7.46,4.00){\scalebox{1.6}{$l$}}
     \rput[cc]{0}(0.40,2.00){\scalebox{1.6}{$N$}}
     \rput[cc]{0}(7.46,0.00){\scalebox{1.6}{$\phi$}}
     \endpspicture
\end{center}

%% file: decay2.tex
 \begin{center}
     \pspicture*(-0.50,-0.5)(8.5,4.5)
     \psset{linecolor=lightgray}
     \qdisk(4,2){1cm}
     \psset{linecolor=black}
     \pscircle[linewidth=1pt](4,2){1cm}
     \psline[linewidth=1pt](4.9,2.5)(7.03,3.75)
     \psline[linewidth=1pt](3.00,2.00)(0.90,2.00)
     \psline[linewidth=1pt,linestyle=dashed](4.9,1.5)(7.03,0.25)
     \psline[linewidth=2pt]{<-}(6.17,3.25)(6.25,3.30)
     \psline[linewidth=2pt]{<-}(6.17,0.75)(6.25,0.70)
     \rput[cc]{0}(7.46,4.00){\scalebox{1.6}{$\bar{l}$}}
     \rput[cc]{0}(0.40,2.00){\scalebox{1.6}{$N$}}
     \rput[cc]{0}(7.46,0.00){\scalebox{1.6}{$\bar{\phi}$}}
     \endpspicture
\end{center}

%% file: decay3.tex
 \begin{center}
     \pspicture*(-0.50,-0.5)(8.5,4.5)
     \psset{linecolor=lightgray}
     \qdisk(4,2){1cm}
     \psset{linecolor=black}
     \pscircle[linewidth=1pt](4,2){1cm}
     \psline[linewidth=1pt](5.0,2.00)(7.10,2.00)
     \psline[linewidth=1pt](3.13,2.50)(0.97,3.75)
     \psline[linewidth=1pt,linestyle=dashed](3.13,1.50)(0.97,0.25)
     \psline[linewidth=2pt]{<-}(1.83,3.25)(1.75,3.30)
     \psline[linewidth=2pt,linestyle=dashed]{<-}(1.83,0.75)(1.75,0.70)
     \rput[cc]{0}(7.60,2.00){\scalebox{1.6}{$N$}}
     \rput[cc]{0}(0.54,4.00){\scalebox{1.6}{$l$}}
     \rput[cc]{0}(0.54,0.00){\scalebox{1.6}{$\phi$}}
     \endpspicture
\end{center}

%% file: decay4.tex
 \begin{center}
     \pspicture*(-0.50,-0.5)(8.5,4.5)
     \psset{linecolor=lightgray}
     \qdisk(4,2){1cm}
     \psset{linecolor=black}
     \pscircle[linewidth=1pt](4,2){1cm}
     \psline[linewidth=1pt](5.0,2.00)(7.10,2.00)
     \psline[linewidth=1pt](3.13,2.50)(0.97,3.75)
     \psline[linewidth=1pt,linestyle=dashed](3.13,1.50)(0.97,0.25)
     \psline[linewidth=2pt]{->}(1.83,3.25)(1.75,3.30)
     \psline[linewidth=2pt,linestyle=dashed]{->}(1.83,0.75)(1.75,0.70)
     \rput[cc]{0}(7.60,2.00){\scalebox{1.6}{$N$}}
     \rput[cc]{0}(0.54,4.00){\scalebox{1.6}{$\bar{l}$}}
     \rput[cc]{0}(0.54,0.00){\scalebox{1.6}{$\bar{\phi}$}}
     \endpspicture
\end{center}

%% file: twotwo.tex
 \begin{center}
     \pspicture*(-0.50,-0.5)(8.5,4.5)
     \psline[linewidth=1pt](4,2)(7.03,3.75)
     \psline[linewidth=1pt](4,2)(0.97,3.75)
     \psline[linewidth=1pt,linestyle=dashed](4,2)(0.97,0.25)
     \psline[linewidth=1pt,linestyle=dashed](4,2)(7.03,0.25)
     \psset{linecolor=lightgray}
     \qdisk(4,2){1cm}
     \psset{linecolor=black}
     \pscircle[linewidth=1pt](4,2){1cm}
     \psline[linewidth=2pt]{->}(6.17,3.25)(6.25,3.30)
     \psline[linewidth=2pt]{<-}(1.83,3.25)(1.75,3.30)
     \psline[linewidth=2pt]{<-}(1.83,0.75)(1.75,0.70)
     \psline[linewidth=2pt]{->}(6.17,0.75)(6.25,0.70)
     \rput[cc]{0}(7.46,4.00){\scalebox{1.6}{$l$}}
     \rput[cc]{0}(0.54,4.00){\scalebox{1.6}{$l$}}
     \rput[cc]{0}(0.54,0.00){\scalebox{1.6}{$\phi$}}
     \rput[cc]{0}(7.46,0.00){\scalebox{1.6}{$\phi$}}
     \endpspicture
\end{center}

%% file: twoanti.tex
 \begin{center}
     \pspicture*(-0.50,-0.5)(8.5,4.5)
     \psline[linewidth=1pt](4,2)(7.03,3.75)
     \psline[linewidth=1pt](4,2)(0.97,3.75)
     \psline[linewidth=1pt,linestyle=dashed](4,2)(0.97,0.25)
     \psline[linewidth=1pt,linestyle=dashed](4,2)(7.03,0.25)
     \psset{linecolor=lightgray}
     \qdisk(4,2){1cm}
     \psset{linecolor=black}
     \pscircle[linewidth=1pt](4,2){1cm}
     \psline[linewidth=2pt]{<-}(6.17,3.25)(6.25,3.30)
     \psline[linewidth=2pt]{<-}(1.83,3.25)(1.75,3.30)
     \psline[linewidth=2pt]{<-}(1.83,0.75)(1.75,0.70)
     \psline[linewidth=2pt]{<-}(6.17,0.75)(6.25,0.70)
     \rput[cc]{0}(7.46,4.00){\scalebox{1.6}{$\bar{l}$}}
     \rput[cc]{0}(0.54,4.00){\scalebox{1.6}{$l$}}
     \rput[cc]{0}(0.54,0.00){\scalebox{1.6}{$\phi$}}
     \rput[cc]{0}(7.46,0.00){\scalebox{1.6}{$\bar{\phi}$}}
     \endpspicture
\end{center}